\documentclass[journal]{IEEEtran}
\usepackage[utf8]{inputenc}
\usepackage{amsmath}
\usepackage{amsthm}
\usepackage{amssymb}
\usepackage{graphicx}
\usepackage{cite}
\usepackage{comment}
\usepackage{epstopdf}
\usepackage{float}
\DeclareMathOperator*{\argmin}{arg\,min}

\usepackage{algorithmicx}
\usepackage{amsfonts}
\usepackage{subcaption}
\usepackage{tikz}

\usepackage[noend]{algpseudocode}
\usepackage[ruled,norelsize,linesnumbered]{algorithm2e}
\usepackage{multicol}
\usepackage{circledsteps}

\makeatletter
\newcommand{\removelatexerror}{\let\@latex@error\@gobble}
\g@addto@macro{\@algocf@init}{\SetKwInOut{Parameter}{Parameters}}
\makeatother

\algdef{SE}[DOWHILE]{Do}{doWhile}{\algorithmicdo}[1]{\algorithmicwhile\ #1}%
\makeatletter
\def\BState{\State\hskip-\ALG@thistlm}

\newlength\myindent
\title{\Large\bf A Mechanical System Inspired Microscopic Traffic Model: Modeling, Analysis, and Validation}
\author{ Mohammad R.~Hajidavalloo,
Zhaojian~Li,
Dong~Chen,
Ali~Louati,
Shuo~Feng,
  Wubing B.~Qin
\thanks{\quad  Mohammad Hajidavalloo, Zhaojian Li, and Dong Chen are with the Department of Mechanical Engineering, Michigan State University, East Lansing, MI 48824, USA.
        Email: {\tt\small \{hajidava,lizhaoj1,chendon9\}@egr.msu.edu}}
\thanks{\quad Ali Louati is with the Department of Information Systems, Prince Sattam Bin Abdulaziz University, 11942 Alkharj, Kingdom of Saudi Arabia.
        Email: {\tt\small a.louati@psau.edu.sa}}
\thanks{\quad Shuo Feng is with the Department of Civil Engineering, University of Michigan, Ann Arbor, MI 48109, USA.
        Email: {\tt\small fshuo@umich.edu}}
\thanks{\quad Wubing Qin is with the Department of Mechanical Engineering, University of Michigan, Ann Arbor, MI 48109, USA.
        Email: {\tt\small wubing@umich.edu}}
        }%

\IEEEoverridecommandlockouts
\begin{document}

\maketitle
\begin{abstract}
In this paper, we develop a mechanical system inspired microscopic traffic model to characterize the longitudinal interaction dynamics among a chain of vehicles. In particular, we extend our prior work on mass-spring-damper-clutch based car-following model between two vehicles to multi-vehicle scenario. This model can naturally capture the driver's tendency to maintain the same speed as the vehicle ahead while keeping a (speed-dependent) desired spacing. It is also capable of characterizing the impact of the following vehicle on the preceding vehicle, which is generally neglected in existing models. A new string stability criterion is defined for the considered 
multi-vehicle dynamics, and stability analysis is performed on the system parameters and time delays. An efficient online parameter identification algorithm, sequential recursive least squares with inverse QR decomposition (SRLS-IQR), is developed to estimate the driving-related model parameters. These real-time estimated parameters can be employed in advanced longitudinal control systems to enable accurate prediction of vehicle trajectories  for improved safety
and fuel efficiency. The proposed model and the parameter identification algorithm are validated on NGSIM, a naturalistic driving dataset, as well as our own connected vehicle driving data. Promising performance is demonstrated. 
\end{abstract}
\section{Introduction}
Rising traffic congestion has become an increasingly frustrating societal problem, especially in large metropolitan areas across the globe. It has led to a variety of issues including great loss in time and money \cite{inrix}, elevated  stress and frustration in drivers \cite{driving_stress}, and intensified air pollution \cite{pollution}. Based on a recent report from INRIX \cite{inrix},  traffic congestion cost U.S. more than \$300 billion dollars and drivers in big cities spent more than 100 hours in congestion in the year of 2017 alone. A number of traffic control technologies have thus been pursued to mitigate the congestion, including ramp metering \cite{metering1,metering2}, dynamic speed limits \cite{variable_speed1,variable_speed2}, vehicle platooning \cite{Platoon1,Platoon2}, and active traffic light control \cite{traffic_light1,traffic_light2,MARL_traffic}. It is worth noting that all those technologies require accurate estimation and prediction of real-time traffic, which creates a critical need to have a good understanding of the traffic dynamics and flow.

Therefore, a multitude of traffic models have been investigated to study traffic characteristics and  flow evolution. These models can generally be classified into two categories: macroscopic and microscopic. Inspired by continuum fluid flow theories,  macroscopic models focus on the study of macroscopic traffic characteristics such as flow speed, traffic density, and traffic volume \cite{Macro}. These models can be further classified, based on different assumptions made,  as kinematic flow models \cite{Kinematic1,Kinematic2,Kinematic3}, dynamic flow models \cite{Dynamic1,Dynamic2}, and lattice hydrodynamic models \cite{Hydro1,Hydro2}. On the other hand, microscopic traffic models deal with local vehicle interactions in terms of relative spacing, speed, and acceleration for individual vehicles. There are two main types of microscopic models: cellular automata (CA) models and car-following (CF) models. The CA models characterize human driving behaviors using stochastic discrete event system, which is capable of modeling lane change behaviors \cite{CA1,CA2} whereas the CF models are concerned with the following vehicle's interaction with its lead vehicle in a single lane setting \cite{1950,GHR}. As the CF models underpin the important design principals of Advanced Driver Assistant Systems (ADAS) such as adaptive cruise control \cite{review}, it will be the focus of this paper.

Many CF models have been developed since the 1950s \cite{1950,GHR,GHR1,GHR2,Helly1,Helly2,review}, among which  the Gazis-Herman-Rothery (GHR) model is arguably the most popular CF model. It was developed in the late 1950s by the General Motors research lab \cite{GHR} with the underlying hypothesis that the instantaneous acceleration of the ego vehicle is directly proportional and inversely proportional, respectively,  to the relative speed and  the relative distance from the lead vehicle, evaluated at time $\tau$ earlier (i.e., delay due to reaction time). Model parameters including the polynomial orders of the speed  and relative distance terms, as well as a gain term, were calibrated using on-road driving data from wire-linked vehicles. A great number of GHR model variants have been developed since then, proposing different ``optimal'' parameter combinations based on  driving data from various experimental setup \cite{GHR1,GHR2,GHR3}. Another class of popular CF models are the Helly models (also known as optimal velocity model), which introduce the idea of desired spacing dependent on speed and/or acceleration as well as explicitly consider an error term \cite{Helly}. Different experimental setups were later proposed and several variants have then been developed based on various experimental datasets \cite{Helly1,Helly2}. Other types of models also exist, including fuzzy logic-based models \cite{FL}, collision avoidance models \cite{CA1,CA2}, and psychophysical models \cite{AP1,AP2}. The readers are referred to \cite{review} for a comprehensive review of the CF models. Despite the large number of existing CF models, it is pointed out in \cite{review} that the available relationships are still not rigorously understood and proven.

\begin{figure*}[!t]
	\centering 
 \includegraphics[width=0.98\textwidth]{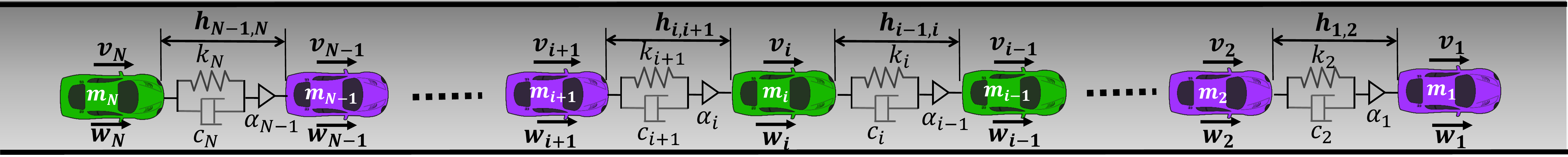}
 \caption{\small A mechanical system inspired traffic model. The car-following behavior of a driver is characterized by a spring, a damper, a force scaling factor, and a clutch (delay).}\label{fig:traffic}
 \vspace{-10pt}
\end{figure*}

To obtain a CF model with better physical interpretability, in our prior work \cite{Li_traffic}, we developed a mass-spring-damper-clutch traffic model that captures many natural driving behaviors in car-following dynamics. Specifically, a mechanical spring between two masses (the lead and ego vehicles) resembles the ego vehicle's tendency to accelerate/decelerate when the relative distance to the preceding vehicle is too large/small; a mechanical damper is used to characterize the ego vehicle's tendency to follow a similar speed as the preceding vehicle, and 
a mechanical clutch system is to model driver's delayed actions due to the reaction time. The model is validated using naturalistic driving data \cite{Li_traffic}. 

Based on our preliminary work \cite{Li_traffic}, in this paper, we extend the mass-spring-damper-clutch model from two vehicles to multiple vehicles. In particular, we model the interactions of multiple vehicles in a single lane as a chain of masses interconnected with springs and dampers between neighboring vehicles. This extended mechanical system-inspired model retains the physical interpretability and can capture the impact of the following vehicle on the leading vehicle, which is neglected in existing CF models. Due to the coupled dynamics resulting from the reaction forces between the masses (representing the vehicles), we define a new string stability criterion and analyze it on system parameters and reaction delays. 

Furthermore, real-time prediction of driving trajectory has shown to be the key enabling technology in ADAS to achieve improved fuel economy and road safety \cite{jin}. Towards this end, we develop an efficient online parameter identification algorithm that exploits inverse QR decomposition  \cite{IQR} to identify the model parameters in real-time. With identified driving-related parameters, the vehicle trajectories can be predicted accordingly. The algorithm is computationally efficient numerically stable, making it suitable for real-time implementations. Furthermore, we validate the proposed model and the parameter identification framework on a real-world driving dataset NGSIM, as well as data from our own connected vehicle studies. Promising performance is demonstrated.

The contributions of this paper include the following. First, we develop a mechanical system inspired microscopic traffic model for multiple vehicle using a chained mass-spring damper system. The new model has great physical interpretability and can characterize the impact of the following vehicle on the preceding vehicle, which is ignored in existing models \cite{GHR,GHR1,Helly,Helly1,CA1}. Secondly, based on the proposed model, we define a new string stability criterion and perform analysis to determine the string stability on different system parameters and various time delays. Last but not least, we develop an online parameter identification algorithm with recursive least square and inverse QR decomposition to estimate the model parameters in real time with great computational efficiency and numerical stability. The proposed models and the online parameter identification algorithm are validated on two naturalistic driving dataset, NGSIM and one collected from our own connected vehicle study, with   promising results demonstrated.

The remainder of this paper is organized as follows. We present our mechanical system inspired traffic model in Section~\ref{sec:2}, followed by the string stability definition and analysis of the model in Section~III. In Section~IV, the online parameter identification algorithm is described whereas the validation of the model and parameter identification framework is presented in Section~V. Finally, concluding remarks and future works are discussed in Section~VI.

\section{Mechanical System Inspired Microscopic Traffic Model}\label{sec:2}
We consider a group of $N$ vehicles on a traffic network link  (in a single lane) where the car-following dynamics is modeled as a mass-spring-damper-clutch system as shown in Figure~\ref{fig:traffic}. Specifically, the vehicles are represented by rigid bodies with springs and dampers connected in-between. The driving behavior of each vehicle $\mathcal V_i,\,i=1,\cdots,N$, is characterized by a tuple $\mathcal V_i=\{m_i, k_i,c_i,\alpha_i,\beta_i\}$, where $m_i$, $k_i$, and $c_i$ are, respectively, the mass, spring stiffness, and damping coefficient to characterize the driving behavior of vehicle $i$.  More specifically, the spring characterizes drivers' tendency to accelerate (or decelerate) when the distance to the preceding vehicle is large (or small), while the damper characterizes that tendency when ego vehicle is slower (or faster) than preceding vehicle. The desired following distance of vehicle $i$ is characterized by the relaxation length of the ``spring'' connecting vehicle $i$ and vehicle $i-1$.    From the observation that the desired following distance, referred to as spacing policy, is typically speed dependent and driver specific \cite{Helly,Helly1}, we characterize  driver $i$'s spacing policy as:
\begin{equation}\label{equ:relaxation}
  X_i^0(v_i) = \begin{cases}
               X_{\min}^0, & \mbox{if } v_i<\underline{v_{i}} \\
               \beta_i\cdot v_i, & \mbox{if }\underline{v_{i}}\leq v_i\leq \bar{v_{i}} \\
               X_{\max}^0, & \mbox{otherwise},
             \end{cases}
\end{equation}
where $\underline{v_{i}}$ and $\bar{v_{i}}$ represent the lower and upper threshold points, respectively; and the slope is denoted as $\beta_i$ that can be interpreted as the desired time headway.

Due to the nature of reaction force from the spring-damper system at the ends, the proposed model can characterize the impact of the following vehicle on the preceding vehicle. For example, if a following tailgates a preceding vehicle, the driver of the preceding vehicle will typically accelerate (or change a lane in a multi-lane situation). This impact is rarely considered in existing literature. Also note that this impact is asymmetric, i.e., the impact of the preceding vehicle on the following vehicle is generally larger than that vice versa. Therefore, we introduce a scaling factor, $\alpha_i\in(0,1]$, to differentiate the two impacts, where $\alpha_i=0$ coincides with the case that the impact of following vehicle on the preceding vehicle is not considered.

With the aforementioned modeling and based on Newton's law, the equations of motion of the vehicles can be written as:
\begin{equation}
\centering
    \begin{aligned}
    \dot{h}_{0,1}(t) &= v_0(t)-v_1(t);\\
  m_1\dot{v}_1  &= k_{1}[h_{0,1}(t-\tau_1)-X_1^0(t-\tau_1)]\\
  &\qquad+c_{1}[v_{0}(t-\tau_1)-v_1(t-\tau_1)]\\
   &\qquad -\alpha_1k_2(h_{1,2}(t-\tau_1)-X_2^0(t-\tau_1))\\
   &\qquad-\alpha_1c_2(v_1(t-\tau_1)-v_2(t-\tau_1));\\
   \dot{h}_{1,2}(t)  &= v_1(t)-v_2(t);\\
    & \qquad\vdots\\
    m_i\dot{v}_i(t) &= k_{i}[h_{i-1,i}(t-\tau_i)-X_i^0(t-\tau_i)]\\
    &\qquad+c_{i}[v_{i-1}(t-\tau_i)-v_i(t-\tau_i)]\\
    &\qquad-\alpha_{i}k_{i+1}[h_{i,i+1}(t-\tau_i)-X_{i+1}^0(t-\tau_i)]\\
    &\qquad-\alpha_i c_{i+1}[v_i(t-\tau_i)-v_{i+1}(t-\tau_i)];\\  
    \dot{h}_{i,i+1}(t) &=v_{i}(t)-v_{i+1}(t);\\
    &\qquad\vdots\\
    \dot{h}_{N-1,N}&=v_{N-1}-v_{N};\\
    m_N\dot{v}_N &= k_{N}[h_{N-1,N}(t-\tau_N)-X_N^0(t-\tau_N)]\\
    &\qquad+c_{N}[v_{N-1}(t-\tau_N)-v_N(t-\tau_N)],
    \end{aligned}
    \label{equ:orisys}
\end{equation}
where $h_{i,i+1}$ represents the distance between vehicle $i$ and vehicle $i+1$, $v_i$ denotes the speed of vehicle $i$, $\tau_i$ represents the delay of vehicle $i$ due to driver $i$'s reaction time, and $X_i^0$ is the equilibrium relaxation length of spring $i$ (desired following distance of vehicle $i$) defined in (\ref{equ:relaxation}).

As mentioned above, the proposed mass-spring-damper-clutch system naturally characterizes a rational driver's car-following behavior. First, a damper, which exerts forces responding to relative speed between two masses, is introduced to characterize the driver's tendency to resist large relative speed (positive or negative). Second, drivers typically follow a desired  relative distance to the lead vehicle, which is captured by a spring that exerts forces responding to deviations from a relaxation equilibrium. Third, the delay induced from driver reaction/vehicle response  is captured by a mechanical clutch that induces time delays for engage and disengage.
Compared to existing microscopic traffic models \cite{GHR,GHR1,GHR2,Helly,Helly1,Helly2}, the proposed  mechanical system inspired model provides interpretable physical insights on the CF dynamics, in contrast to existing models that are mainly derived from data regression. Furthermore, the model is capable of characterizing the impact of the following vehicle on the preceding vehicle, which is neglected in existing models. This impact is frequently observed in daily driving and issues arise when neglecting it in cascading CF models into a chain  to characterize the dynamics of multiple vehicles. 

\vspace{-8pt}
\section{stability analysis}
\begin{figure}
  \includegraphics[width=\linewidth]{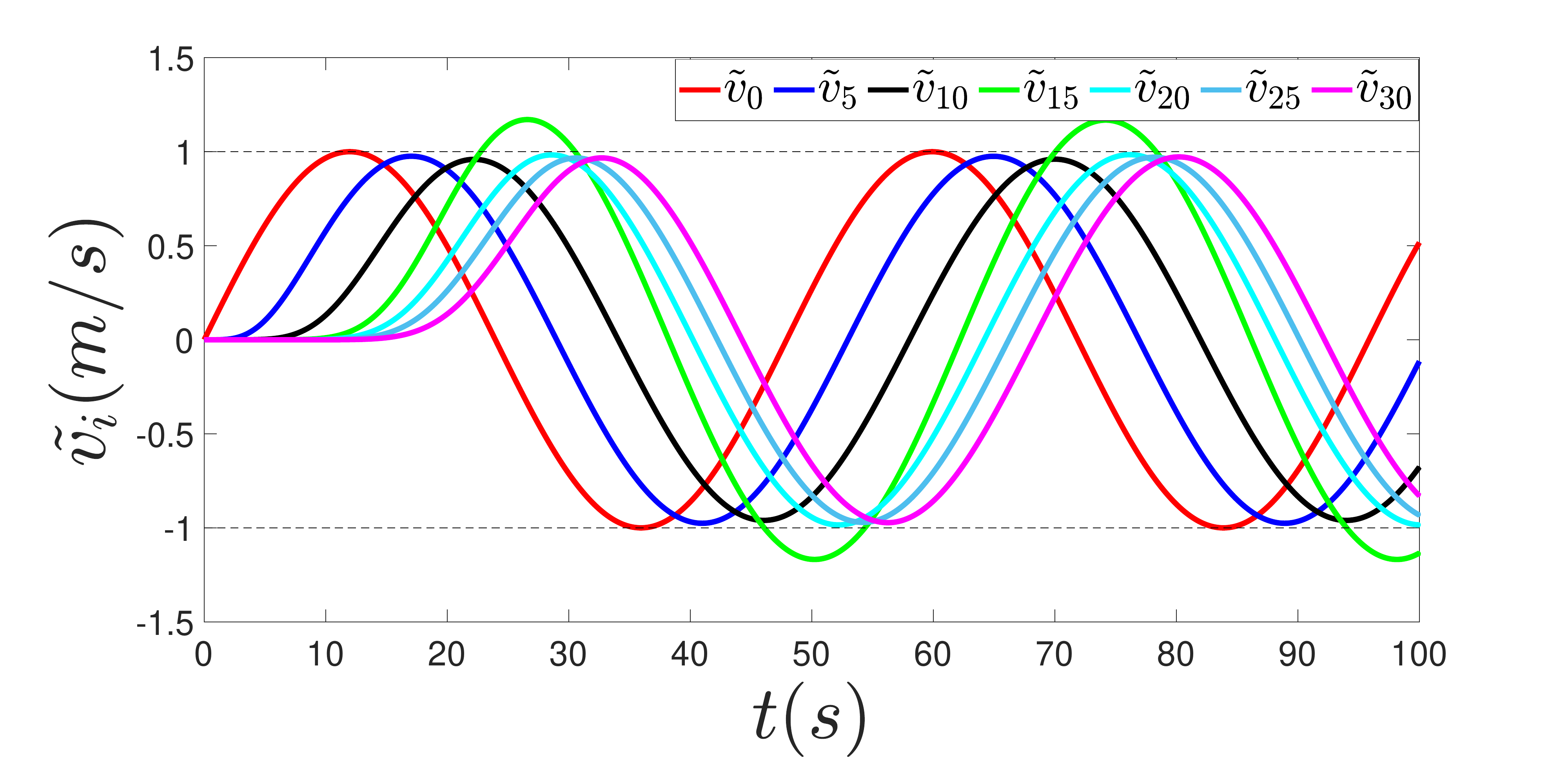}
\caption{\small String unstable vehicle (vehicle 15) in-between the chain of string stable vehicles}
\label{fig:inbetween}
\vspace{-10pt}
\end{figure}

In this section, stability analysis of the above mechanical system-inspired traffic model is presented. From the equations of motion in Eqn.~\ref{equ:orisys}, it is easy to show that the system has a uniform-flow equilibrium at which all following vehicles achieve the same speed as the lead vehicle while  maintaining the desired following distance based on its speed and desired time headway, i.e., 
\begin{equation}
\begin{aligned}
    & v_{0}^{*}=v_{1}^{*}=....=v_{N}^{*}, \\ 
& h_{i-1,i}^{*}={{\beta }_{i}}v_{i}^{*},\\
\end{aligned}
\end{equation}
for $i=1,2,...,N$, where $v_0^*$ is the equilibrium speed of the ``ghost vehicle''. Here the ghost vehicle is defined as the vehicle, physically or virtually, leading the first vehicle and is not part of the studied fleet. To study stability, we define the perturbation on the ghost vehicle as:
\begin{equation}
    \tilde{v}_{0}=  v_{0} - v_{0}^{*},
\end{equation}
which will result in fluctuations in the whole system with the following defined perturbations
\begin{equation}
\begin{aligned}
     \tilde{v}_{i} &= v_{i} -v_{i}^{*},\\
\tilde{h}_{i-1, i} &=h_{i-1, i} -h_{i-1, i}^{*},
\end{aligned}
\end{equation}
for $i=1,2,...,N$. We can then obtain the perturbation dynamics from Eqn.~\ref{equ:orisys} as follows:
\begin{equation}
\centering
    \begin{aligned}
    \dot{\tilde{h}}_{0,1}(t) &= \tilde{v}_0(t)-\tilde{v}_1(t);\\
   m_1\dot{\tilde{v}}_1 &= k_{1}[\tilde{h}_{0,1}(t-\tau_1)-\beta_1\tilde{v}_1(t-\tau_1)]\\
   &+c_{1}[\tilde{v}_{0}(t-\tau_1)-\tilde{v}_1(t-\tau_1)]\\
   & -\alpha_1k_2(\tilde{h}_{1,2}(t-\tau_1)-\beta_2\tilde{v}_2(t-\tau_1))\\
   &-\alpha_1c_2(\tilde{v}_1(t-\tau_1)-\tilde{v}_2(t-\tau_1));\\
    \dot{\tilde{h}}_{1,2}(t) &= \tilde{v}_1(t)-\tilde{v}_2(t);\\
    & \vdots\\
    \dot{\tilde{h}}_{N-1,N}&=\tilde{v}_{N-1}-\tilde{v}_{N};\\
    m_N\dot{\tilde{v}}_N &= k_{N}[\tilde{h}_{N-1,N}(t-\tau_N)-\beta_N \tilde{v}_N(t-\tau_N)]+\\
    &c_{N}[\tilde{v}_{N-1}(t-\tau_N)-\tilde{v}_N(t-\tau_N)],
    \end{aligned}
    \label{equ:orisys2}
\end{equation}
We define, respectively, the perturbed state vector, the disturbance input, and the output vector as:
\begin{equation}
\begin{aligned}
  x&={{\left[ {{{\tilde{h}}}_{0,1}},{{{\tilde{v}}}_{1}},\cdots ,{{{\tilde{h}}}_{N-1,N}},{{{\tilde{v}}}_{N}} \right]}^{\text{T}}},\\
  u&={{{\tilde{v}}}_{0}}, \qquad y={{[{{{\tilde{v}}}_{1}},{{{\tilde{v}}}_{2}},...,{{{\tilde{v}}}_{N}}]}^{T}},
 \end{aligned}
\end{equation}
which allows us to write Eqn.~\ref{equ:orisys2} in the following compact form:
\begin{equation}
    \begin{aligned}
  \dot{x}(t)&=A_0x(t)+\sum\limits_{i=1}^{N}{{{A}_{i,d}}x(t-{{\tau }_{i}})}+Bu(t-{{\tau }_{1}}) \\ 
  y(t)&=Cx(t), \\ 
\end{aligned}
\label{equ:orisys3}
\end{equation}
where the matrices $A_0$, $A_{i,d}$, $B$ and $C$ can be found by inspecting Eqn.~\ref{equ:orisys2}. The plant and string stability  of the above linear time-delay system can be analyzed by either approximating the system with an equivalent non-delay linear system (i.e., to find equivalent $A_e$, $B_e$ and $C_e$ matrices in a higher dimension) or using iterative approaches to directly analyze the time-delay systems, see e.g., \cite{timed1,timed2}. We adopt the former approach here as it can be easily done via numerical software (e.g., with the \textit{pade} command in Matlab). Given the system matrices $A_0$, $A_{i,d}$, $B$, $C$ and the delays $\tau_i,\,i=1,\cdots,N$, the plant stability criterion reduces to checking whether all eigenvalues of the equivalent  matrix $A_e$ have negative real parts.   
We next examine  the string stability of the considered time-delay system in (\ref{equ:orisys3}). The traditional definition of string stability for a platoon of vehicles is that the spacing error between vehicles\cite{Rajamani2012} (i.e., the difference between the desired spacing and the actual spacing) or velocity fluctuations \cite{Qin2019_2}  attenuates along  the string of vehicles, which will guarantee that the disturbance impact decreases from vehicle to vehicle to achieve stability. However, this stability criterion does not apply to our new traffic model, which considers the impact of the following vehicle on the preceding vehicle that is not modeled in existing models, i.e., the impact is bi-directional and coupled. Therefore, the new definition of string stability must require the attenuation of disturbances on all vehicles. This more strict definition is necessary for our new traffic model as the disturbance wave propagates both forwards and backwards due to the bi-directional impacts between lead and following vehicles, and any vehicle in-between can oscillate out of bounds. As a motivating example, we find a combination of the parameters ${k}_{i}$, ${c}_{i}$ and simulate a chain of 30 vehicles whose results are illustrated in Fig.~\ref{fig:inbetween}.  It can be seen that some vehicles (e.g., vehicle 15) in-between the chain amplify the incurred disturbance on the ghost vehicle whereas other vehicles attenuate the disturbance. This clearly shows the need of the new definition where the attenuation of the disturbance on the ghost vehicle for all vehicles across a spectrum of frequencies must be satisfied. Otherwise, based on traditional definitions of string stability (e.g., \cite{Qin2019_2}), the chain of vehicles will be incorrectly considered as string stable.

Therefore, we determine the string stability  by examining the frequency response from $\tilde{v}_0$ to $\tilde{v}_i$, $i=1,\cdots,N$, and the string stability criterion for our model is defined as:
\begin{equation}
    \underset{\omega }{\mathop{\sup }}\,\left| \frac{{{{\tilde{V}}}_{i}}(j\omega )}{{{{\tilde{V}}}_{0}}(j\omega )} \right|\le 1,\text{   }\forall i=1,2,...,N,
\end{equation}
where $\tilde{V_0}(s)$ and $\tilde{V_i}(s)$ are the Laplace transform of $\tilde{v_0}$ and $\tilde{v_i}$, respectively. Note that the Laplace transform of (\ref{equ:orisys3}) is
\begin{equation}
    G(s)=C(sI-{{A}_{0}}-\sum\limits_{i=1}^{N}{{{A}_{i,d}}{{e}^{-s{{\tau }_{i}}}}{{)}^{-1}}B}{{e}^{-s{{\tau }_{1}}}},
\end{equation}
where the $i$-th element is
\begin{equation}
    G_i(s)=\frac{{\tilde{V}_{i}}(s)}{{\tilde{V}_{0}}(s)}.
\end{equation}

While the above plant and string stability criteria can be applied to an arbitrary set of system parameters and time delays, to better showcase the analysis we consider uniform parameters across the vehicle fleet and check the stability by varying the parameters in the 2D space of  ${{\tilde{k}}_{i}}=\frac{{{k}_{i}}}{{{m}_{i}}}$ and ${{\tilde{c}_{i}}=\frac{{{c}_{i}}}{{{m}_{i}}}}$ while keeping other parameters constant, with $m_i=1$, $\beta_i=1$, $\alpha_i=0.2$ for $i=1,2,...,N$ where $N=30$. 

\begin{figure}[!t]
   \includegraphics[width=\linewidth]{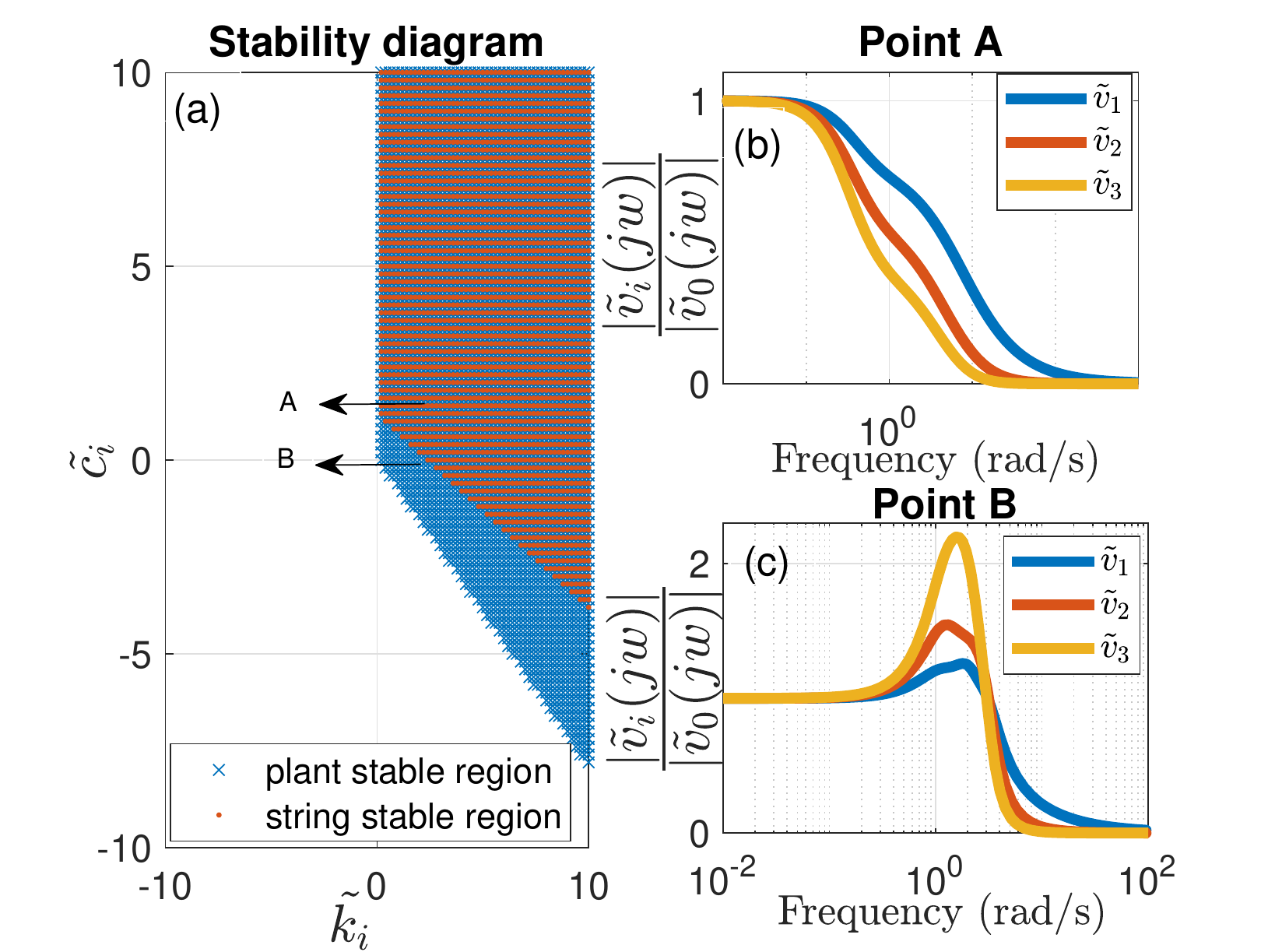}
\caption{\small (a) Plant stable and string stable regions for the no-delay case. Point A represents a string stable set of parameters and point B represents a string unstable set of parameters in the parameter space; (b) Gain plot of the frequency response for the first three vehicles for point A; and (c) Gain plot of the frequency response for the first three vehicles for point B.}
\label{fig:stabdiag_noDelay}
\vspace{-5pt}
\end{figure}

We first consider the case that no delays are present in the system. For each pair $(\tilde{k}_{i},{\tilde{c}}_{i})$ in the grid space of $-10 <\{{\tilde{k}_{i}},{{\tilde{c}}_{i}}\}<10$ with a grid size of 0.1, the aforementioned plant stability and string stability analyses are performed, whose results are shown in Fig.~\ref{fig:stabdiag_noDelay}(a). Note that plant stability is the perquisite for string stability. Therefore, we only look for string stable regions inside plant stable regions. 


 \begin{figure}[!t]
   \vspace{-10pt}
  \includegraphics[width=0.98\linewidth]{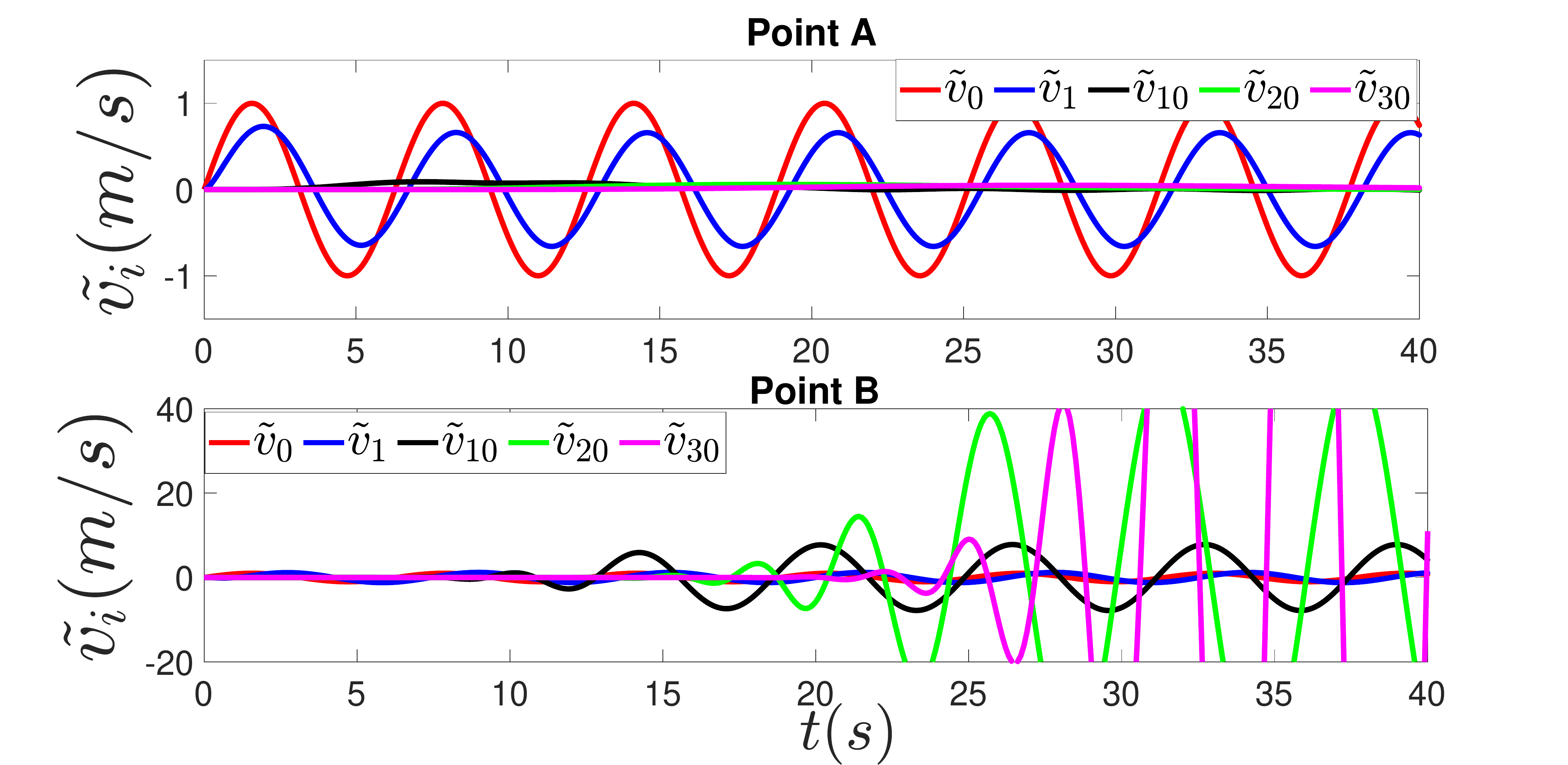}
\caption{\small Speed perturbations of the ghost vehicle and 4 sample vehicles in the no-delay case with input frequency of 1 $rad/s$: (a) for parameter set A, the perturbations are attenuated; and (b) for parameter set B, the perturbations are amplified.}
\label{fig:stabdiag(b)}
\vspace{-10pt}
\end{figure}

A string stable parameter pair (point A) and a plant stable but not string stable parameter pair (point B) are further examined, where the frequency responses for four sample vehicles are shown in Fig.~\ref{fig:stabdiag_noDelay}(b) and Fig.~\ref{fig:stabdiag_noDelay}(c), respectively. It can be seen that the parameter pair corresponding to point A has an attenuating gain (i.e., $\leq1$) for all three vehicles across all frequencies while the parameter pair corresponding to point B has amplifying gains in some frequencies. The vehicle speed perturbations of the three vehicles and the ghost vehicle are shown in Fig.~\ref{fig:stabdiag(b)}, where we add some sinusoidal fluctuations on the ghost vehicle, i.e.,
\begin{equation}
    {{\tilde{v}}_{0}}(t)={{v}_{0}}^{\text{amp}}\sin (\omega t).
\end{equation}
As the underlying dynamics (\ref{equ:orisys2}) is linear, the sinusoidal fluctuation in the ghost vehicle leads to sinusoidal fluctuations in following vehicles as indicated by Fig.~ \ref{fig:stabdiag(b)}. This figure also shows that the speed perturbations of the illustrated following vehicles are attenuated under the parameter set A whereas the vehicle perturbations are amplified under the parameter set B.

\begin{figure}[!h]
   \vspace{-10pt}
  \includegraphics[width=\linewidth]{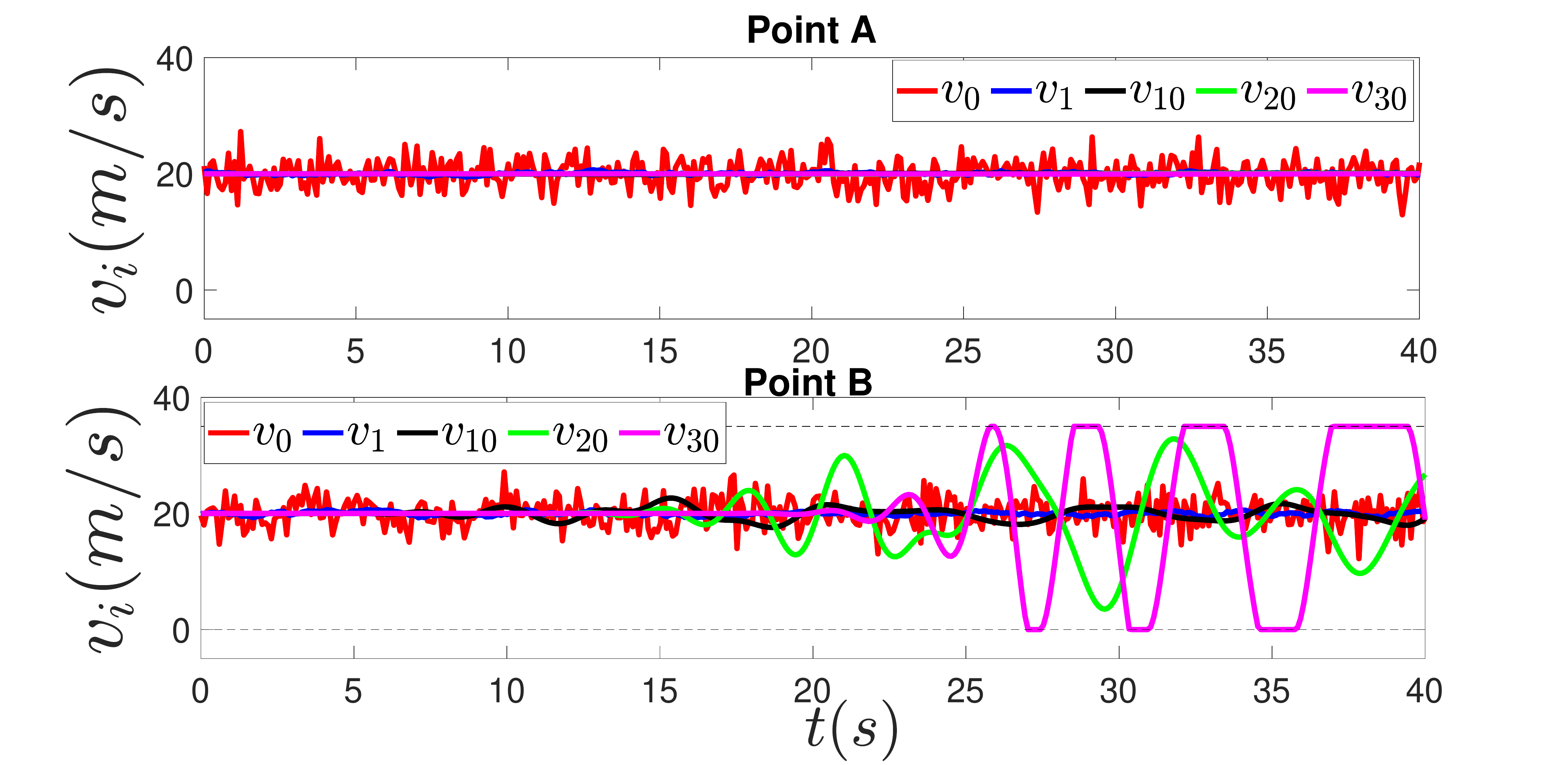}
\caption{\small Speed profiles of the ghost vehicle and 4 sample vehicles in the no-delay case with white noise perturbations: (a) for parameter set A, the perturbations are attenuated; and (b) for parameter set B, the perturbations are amplified and cause stop-and-go traffic jam.}
\label{fig:trrafic2}
\vspace{-10pt}
\end{figure}

In real world scenarios, speed disturbances can contain a broad spectrum of frequencies, and by definition string unstable systems will amplify fluctuations at certain frequencies. If the vehicle chain is long, these amplified fluctuations  will eventually lead to stop-and-go traffic jam (also referred to as ghost traffic jam or phantom traffic jam), which is shown in Fig.~\ref{fig:trrafic2} that characterizes the response of white-noise disturbance on the ghost vehicle with $v_{i}^{*}=20\,m/s$ for string stable and string unstable cases. For the string unstable case (point B), the fluctuation on the ghost vehicle amplifies along the chain of vehicles, and those velocity profiles demonstrate the stop-and-go traffic jam phenomena (see e.g., vehicle 30). In contrast, for string stable case (point A), the white noise fluctuations of the lead vehicle leads to attenuated fluctuations in the following vehicles. 

\begin{figure}[!h]
\vspace{-10pt}
 \includegraphics[width=0.95\linewidth]{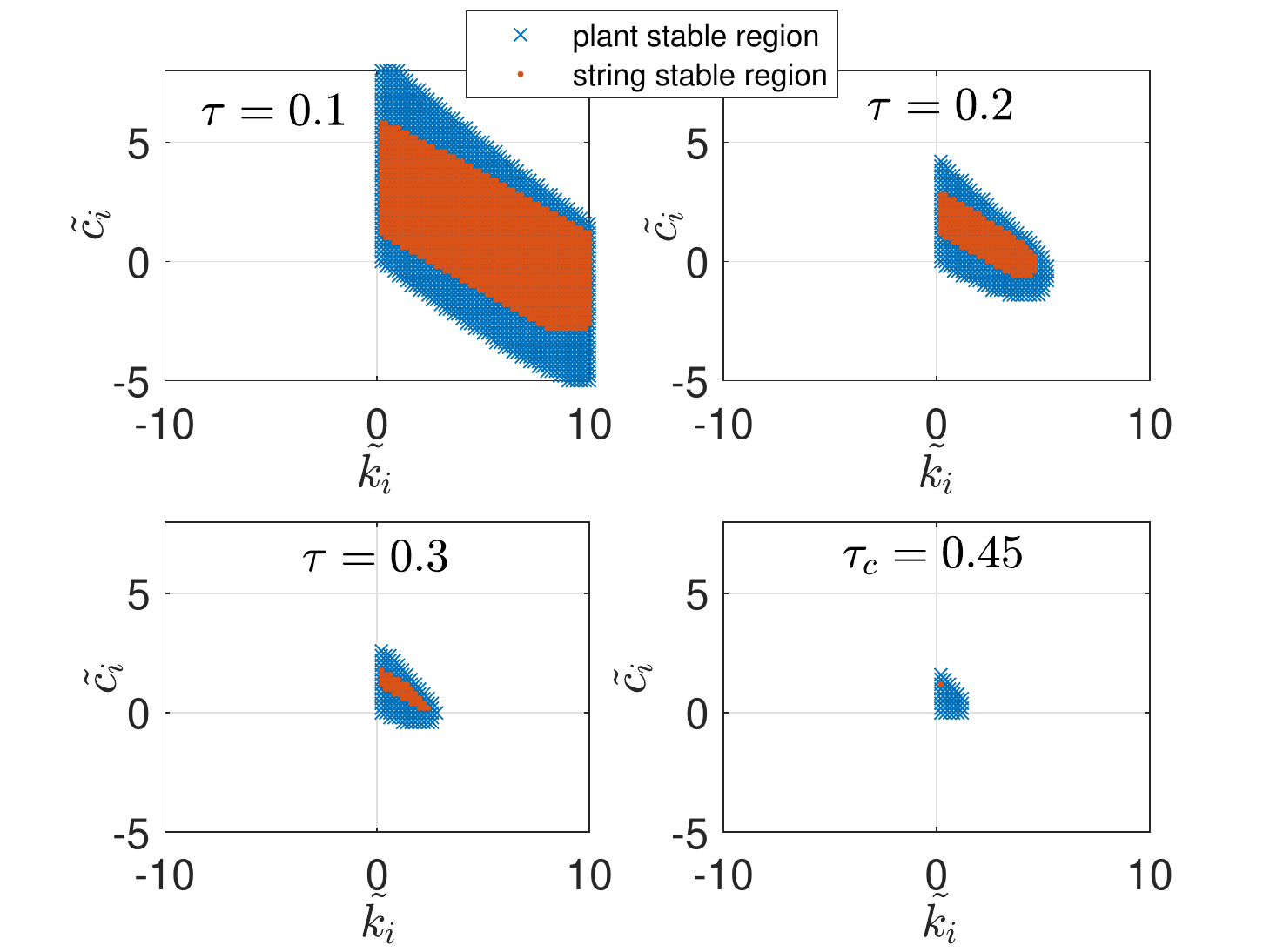}
\caption{\small Plant and string stable regions for the uniform delay case with different delays. }
\vspace{-10pt}
\label{fig:stab_uni_delay}
\end{figure}

We next analyze plant and string stability with time delays, which correspond to the reaction time of human drivers and/or the intermittent vehicle-to-vehicle (V2V) communication delays \cite{Qin2019_3}. We first consider a fleet of $30$ vehicles with a uniform delay across all the vehicles, i.e. $\tau_1=\tau_2=...=\tau_N=\tau$. The stability results for the uniform delays of $\tau=0.1,\,0.2,\,0.3,\,0.45$ are shown in Fig.~\ref{fig:stab_uni_delay}. It is clear (and as expected) that as the delay increases, the plant stable and string stable regions shrink. When the delay increases to a critical value (e.g., 0.45s in our simulation), there is almost no parameter set that can make the system string stable.  


\begin{figure}[t]
 \includegraphics[width=\linewidth]{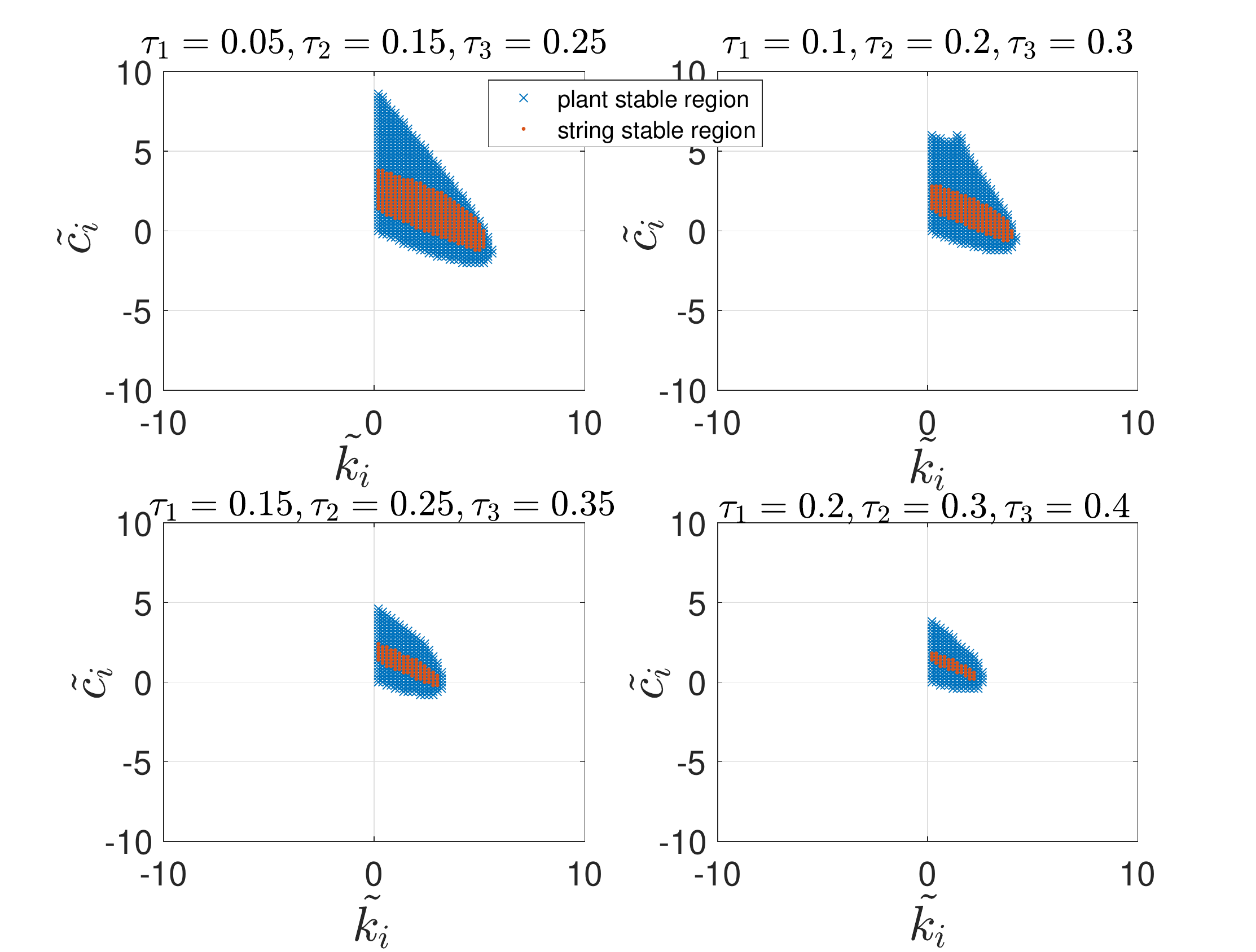}
\caption{\small Plant and string stable regions for the non-uniform delay case with different delay values. Point A represents a string stable set of parameters and point B represents a string unstable set of parameters in the parameter space.}
\label{fig:stab_non_uni_delays}
\end{figure}

We next investigate the case of non-uniform delays, which is a more general and realistic scenario that incorporates vehicle/driver heterogeneity. Due to the large number of delay combinations with multiple vehicles, we only show the case with three following vehicles. The stability region of four different delay combinations are shown in Fig.~\ref{fig:stab_non_uni_delays}, which also shows a reasonable trend that   as delays increase for each vehicle, the plant stable and string stable regions shrink. 

\section{Online Parameter Identification}
Speed prediction plays a critical role in advanced longitudinal controls to achieve improved safety and fuel efficiency  \cite{CCC}. It is generally assumed in existing cooperative cruise control frameworks that the dynamics/control policies of preceding vehicles are available \cite{CCC1,CCC2}. While this is a reasonable assumption in a platoon of fully automated vehicles, it is difficult to achieve when human drivers are involved in the platoon. Therefore, there is a crucial need to predict the maneuvers of human drivers such that a ``comprehensive preview'' can be provided. As the human driver behavior is characterized by the parameters in the proposed mechanical system inspired model, in this section we develop an efficient online parameter identification framework to enable real-time identification of those driving-related parameters and thus subsequently predict its future speed online.

More specifically, we consider a centralized estimation scheme where all participating vehicles share its vehicle data (e.g., position, speed) to a central agent through wireless communication. This can be achieved through vehicle-to-infrastructure or vehicle-to-edge cloud communications. To identify the parameters with received measurements, we first discretize the system in (\ref{equ:orisys}) using the Euler's method with sampling time $\Delta$, which is reduced to
\begin{equation}
\centering
    \begin{aligned}
  m_1 &(v_1(k)-v_1(k-1))/{\Delta}  = k_{1}[h_{0,1}(k-d_1)-X_1^0(k-d_1)]\\
  &\qquad+c_{1}[v_{0}(k-d_1)-v_1(k-d_1)]\\
   &\qquad -\alpha_1k_2(h_{1,2}(k-d_1)-X_2^0(k-d_1))\\
   &\qquad-\alpha_1c_2(v_1(k-d_1)-v_2(k-d_1));\\
    &\qquad\qquad\qquad\qquad\vdots\\
    m_N &(v_N(k)-v_N(k-1))/{\Delta} = k_{N}[h_{N-1,N}(k-d_N)\\
    &\quad-X_N^0(k-d_N)]+c_{N}[v_{N-1}(k-d_N)-v_N(k-d_N)],
    \end{aligned}
    \label{equ:dissys}
\end{equation}
where $d_i=round(\tau_i/\Delta)$ is the discretized delay steps. Note that as the relative speed $v_{i-1}-v_i$ and the relative distance  $h_{i-1,i}$ are measured, (\ref{equ:dissys}) can be written as $z(k)=f(\mathcal{D}(k);\theta)$, where $z(k)=[(v_1(k)-v_1(k-1))/\Delta,\cdots,(v_N(k)-v_N(k-1))/\Delta]^\text{T}$ is the measured output; $\mathcal{D}(k)=[h_{0,1}(k),\,v_0(k),\,v_1(k),\cdots,h_{N-1,N}(k),\,v_{N-1}(k),\,v_{N}(k)]$ is the vector of available measurements; and $\bar\theta=[m_1,k_1,c_1,\alpha_1,\beta_1,\cdots]$ is a vector of parameters to be estimated. Note that the function $f$ is nonlinear in the parameter vector $\bar\theta$ as the parameters are nonlinearly coupled. To facilitate the parameter identification, we make some simplifications here by assuming that the mass terms $m_i$, the desired time headways $\beta_i$, and the scaling factors $\alpha_i$ are homogeneous and constant across the vehicles. As a result, (\ref{equ:dissys}) can be further reduced to the following linear form:
\begin{equation}
    z(k)=\Phi(k)\theta,
\end{equation}
where $z(k)\in\mathbb{R}^n$ is the discretized acceleration vector same as above; $\Phi(k)\in\mathbb{R}^{n\times 2n}$ is the data matrix derived from (\ref{equ:dissys}); and  $\theta=[k_1,c_1,\cdots,k_N,c_N]^\text{T}\in\mathbb{R}^{2n}$ is the parameter vector to be estimated. Given a past data sequence $(\Phi(k),z(k)),\,k=1,2,\cdots,K$, the least-square solution of the optimal parameter can be calculated as:
\begin{equation}
\theta^*=\argmin_{\theta}\sum_{k=1}^K \|\Phi(k)\theta-z(k)\|^2,
\end{equation}
or equivalently
\begin{equation}\label{equ:OLS}
\theta^*=\argmin_{\theta}\|\Psi(K)\theta-Z(K)\|^2,
\end{equation}
where $\Psi(K)=[\Phi(1);\cdots;\Phi(K)]$ and $Z(K)=[z(1);\cdots;z(K)]$ are the stacked data matrices.

It is straightforward to show that the solution to the optimization problem (\ref{equ:OLS}) is:
\begin{equation}
    \theta^*=(\Psi(K)^\text{T}\Psi(K))^{-1}\Psi(K)^\text{T}Z(K).    
\end{equation}
Note that as the time step number $K$ increases, the sizes of data matrices $\Psi(K)$ and $Z(K)$ grow dramatically, making it infeasible for online implementation. Therefore, we develop a recursive algorithm --  sequential recursive least squares with inverse QR-decomposition (SRLS-IQR) -- to enable parameter update in real time. In particular, at each time step $k$, we use data pairs ($\Phi_i(k),z_i(k)$) from each row $i$, $i=1,\ldots,2n$, of the data matrix $\Phi(k)$ and output $z(k)$ to update the parameter vector $\theta$ sequentially (for a total of $2n$ times). Furthermore, we exploit inverse QR decomposition for the recursive least square update due to its great numerical stability and computational efficiency \cite{IQR}. The SRLS-IQR algorithm is detailed in Algorithm~\ref{algo:1}.

\begin{figure}[!t]
\removelatexerror
\scalebox{0.9}{
\begin{algorithm}[H]
\SetAlFnt{\small}
    \SetKwInOut{Input}{Input}
    \SetKwInOut{Output}{Output}
\caption{Sequential Recursive Least Squares with Inverse QR Decomposition}
\label{algo:1}
\SetAlgoLined
\Parameter{ RLS forgetting factor $\lambda$, inverse matrix initialization parameter $\delta$.}
\Input{$\{\Phi(k),z(k)\}_{k=1}^K$.}
\Output{$\{\theta(k)\}_{k=1}^{K}$.}
\vspace{0.2em}
\hrule
\vspace{0.2em}
{\bf initialize} $\theta(0) \gets \mathbf 0_{2n\times1}$, $R^{-\text T} \gets \delta \mathbf I_{2n}$;\\
\For{$k=1 \to K$}{
    \tcc{ \it Sequential update for each row of $\Phi(k)$}
    \For{$l=1\to 2n$}{
        {\bf set} {\small{$x^\text{T}(k)\gets \text{row } l \text{ of } \Phi(k)\;, y^\text{T}(k)\gets \text{row } l \text{ of } z(k)$;}}\\

        \tcc{\it Parameter update}
        {\bf initialize} $u_{j,m}\gets 0, b_0\gets 1, 1\leq j\leq 2n, \, m<j;$\\

        \For{$i=1\to 2n$}{
        $a_i=\lambda^{-1/2}\sum_{j=1}^ir_{ij}(d,k-1)x_k(j)$;\\
        $b_i=\sqrt{b_{i-1}^2+a_i^2}$;\\
        $s_i=a_i/b_i;$\\
        $c_i=b_{i-1}/b_i;$\\
            \For{$j=1\to i$}{
                $r_{ij}(k)=\lambda^{-1/2}c_ir_{ij}(k-1)-s_iu_{i-1,j}$;\\
                $u_{i,j}=c_iu_{i-1,j}+\lambda^{-1/2}s_ir_{ij}(k-1)$;\\
            }

        }
        $o(k)=e(k|k-1)/b_{2n}$;\\
        \For{$i=1\to 2n$}{
        $\theta(k)\gets \theta(k-1)+o(k)u_{i,2n}$;\\
        }
}
}
\end{algorithm}
}
\end{figure}

\section{Naturalistic driving data validation}
In this section, we validate the proposed mechanical system-inspired traffic model as well as the SRLS-IQR online parameter identification on two naturalistic driving datasets: NGSIM \cite{alexiadis2004next} and our own data collected using connected vehicles. 
\vspace{-12pt}
\subsection{NGSIM dataset validation}
\begin{figure}[!h]
\vspace{-8pt}
\centering
 \includegraphics[width=0.85\linewidth]{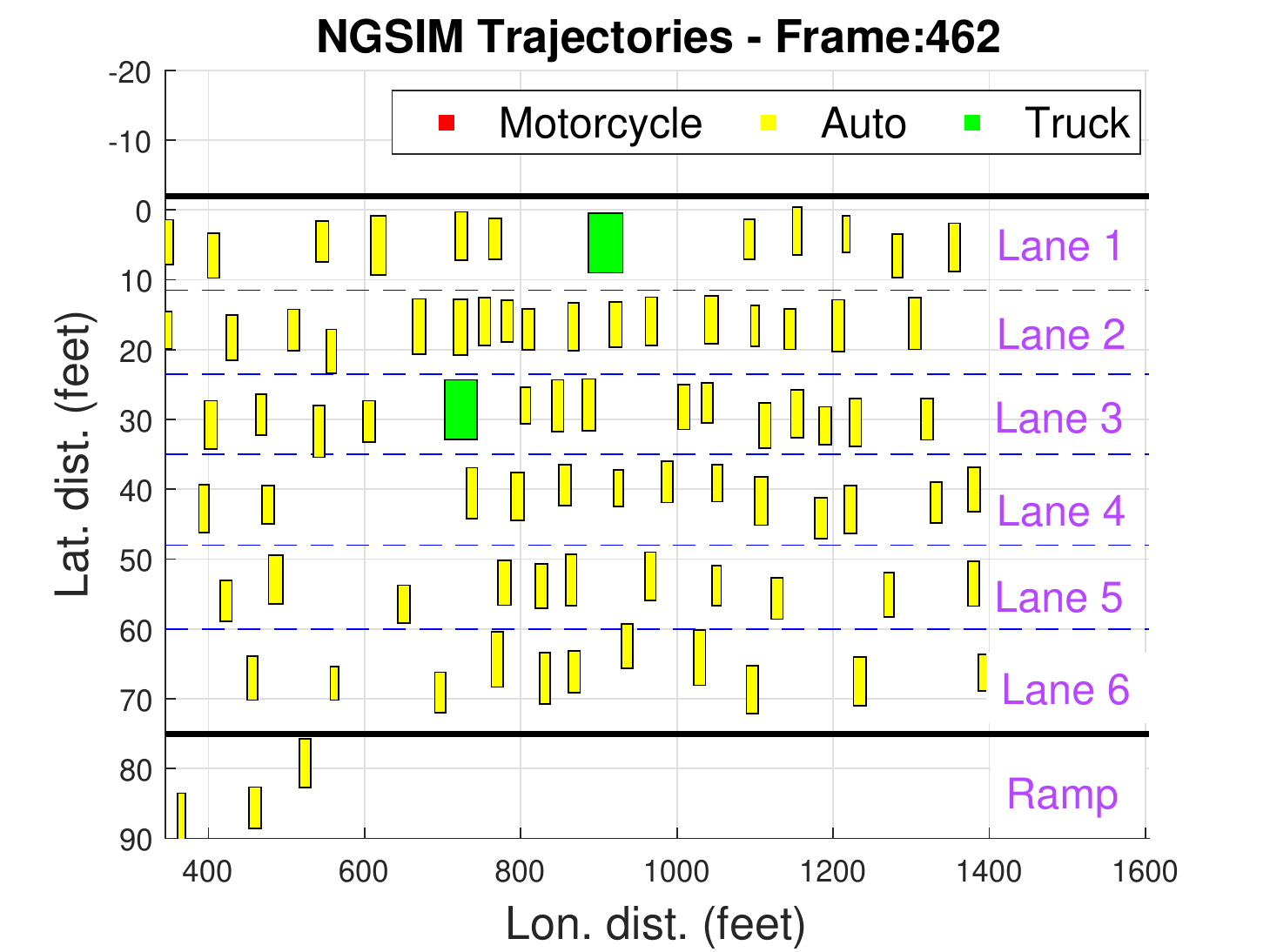}
 \caption{\small A snapshot of the vehicle positions from the NGSIM dataset (video frame 462).}\label{fig:NGSIM}
 \vspace{-2pt}
\end{figure}

In this subsection, we evaluate the proposed framework using  real-world vehicle trajectory data on US Interstate 80 (I-80 San Francisco) collected by the NGSIM program \cite{alexiadis2004next}. A total of 45-minutes vehicle data on a 500-meter long, six-lane study area was recorded by eight synchronized digital video cameras. The video was then transcribed into vehicle trajectory data (e.g., position, relative position, speed) at a 10 Hz frequency. Note that the vehicle trajectory data involves multiple lanes and some lane change trajectories were recorded. Therefore, in order to  validate our proposed traffic model 
which is based on a single lane, we process the dataset by filtering out  lane-changing scenarios and only keep the time periods with no lane changes. We focus on the sub-area of the 400-1600 feet road segment (see Fig.~\ref{fig:NGSIM} for a snapshot of the vehicle locations on the considered road segment).  A total of 8 single-lane vehicle trajectory sets are produced for validation. 

With the processed data, we then run the SRLS-IQR algorithm (Algorithm 1) to identify the driving related parameters (i.e., spring stiffness $k$ and damping coefficient $c$) with the following hyper-parameters that are shown in Table~\ref{tab:1}. 
\begin{table}[!h]
\vspace{-5pt}
\caption{Hyper-parameters used in the SRLS-IQR algorithm.}
\label{tab:1}
\centering
\begin{tabular}{|c|c|c|}
\hline
\hline
 Mass ($M $) &Scaling factor ($\alpha$) &Time headway ($\beta$) \\
 \hline
1 $[kg]$&0.1 &2.5 $[s]$\\
 \hline
 Forgetting factor ($\lambda $) &Matrix init. param. ($\delta$) &Delay steps ($d$) \\
 \hline
0.95 &100 &5 \\
 \hline 
 \hline
\end{tabular}
\vspace{-13pt}
\end{table}

\begin{figure}[!t]
\vspace{0pt}
\centering
 \includegraphics[width=0.8\linewidth]{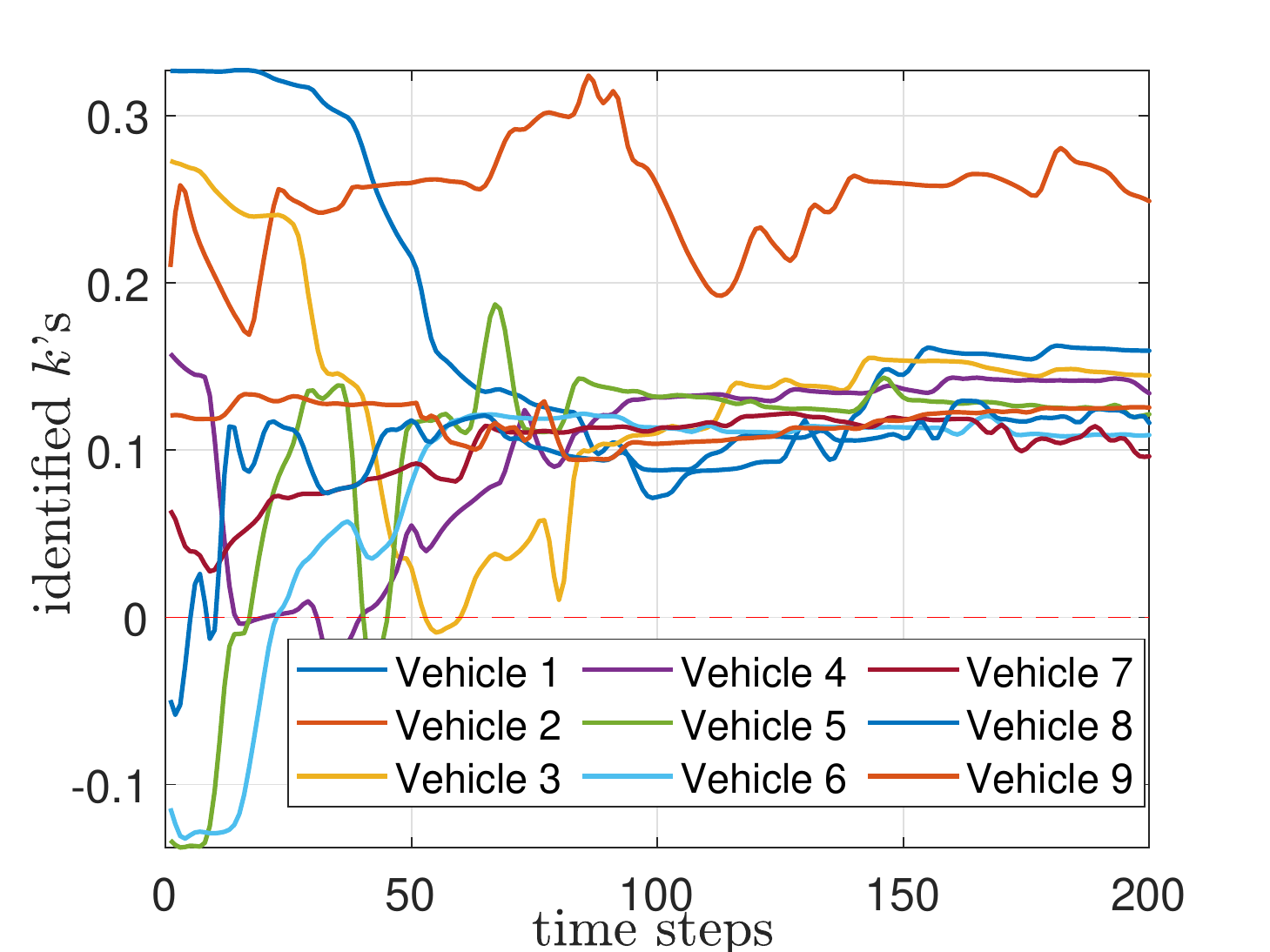}
 \caption{\small Real-time updated parameter $k$'s (spring stiffness) of the 9 vehicles in one processed dataset.}\label{fig:paraK}
 \vspace{-10pt}
\end{figure}

\begin{figure}[!h]
\vspace{0pt}
\centering
 \includegraphics[width=0.8\linewidth]{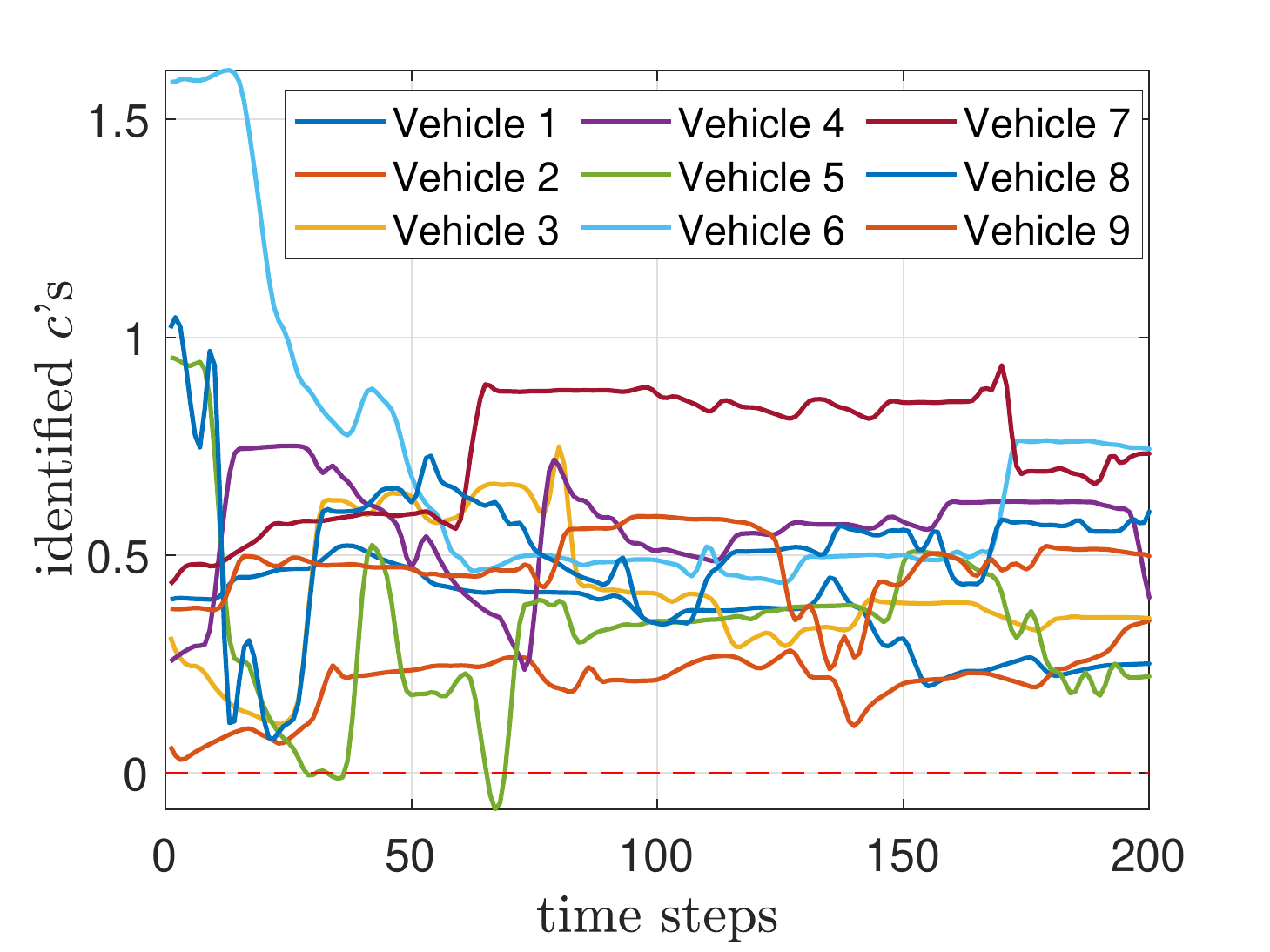}
 \caption{\small Real-time updated parameter $c$'s (damping coefficient) of the 9 vehicles in one processed dataset.}\label{fig:paraC}
\end{figure}

The identified parameters of a trajectory with 9 vehicles (Episode 1) are shown in Fig.~\ref{fig:paraK} and Fig.~\ref{fig:paraC}. Several observations are made here. First, with some initial learning (e.g., after $\sim$100 steps or 10 seconds), the identified parameters are relatively stable with small variations. Second, the identified parameters ($k$'s and $c$'s) are positive for most of the time, which is consistent with the physical interpretation as spring stiffness and damping coefficient, respectively, that are non-negative. Third, the parameters for different drivers are at different levels, which is sensible as different drivers have different driving patterns. 

\begin{figure}[!h]
\centering
 \includegraphics[width=\linewidth]{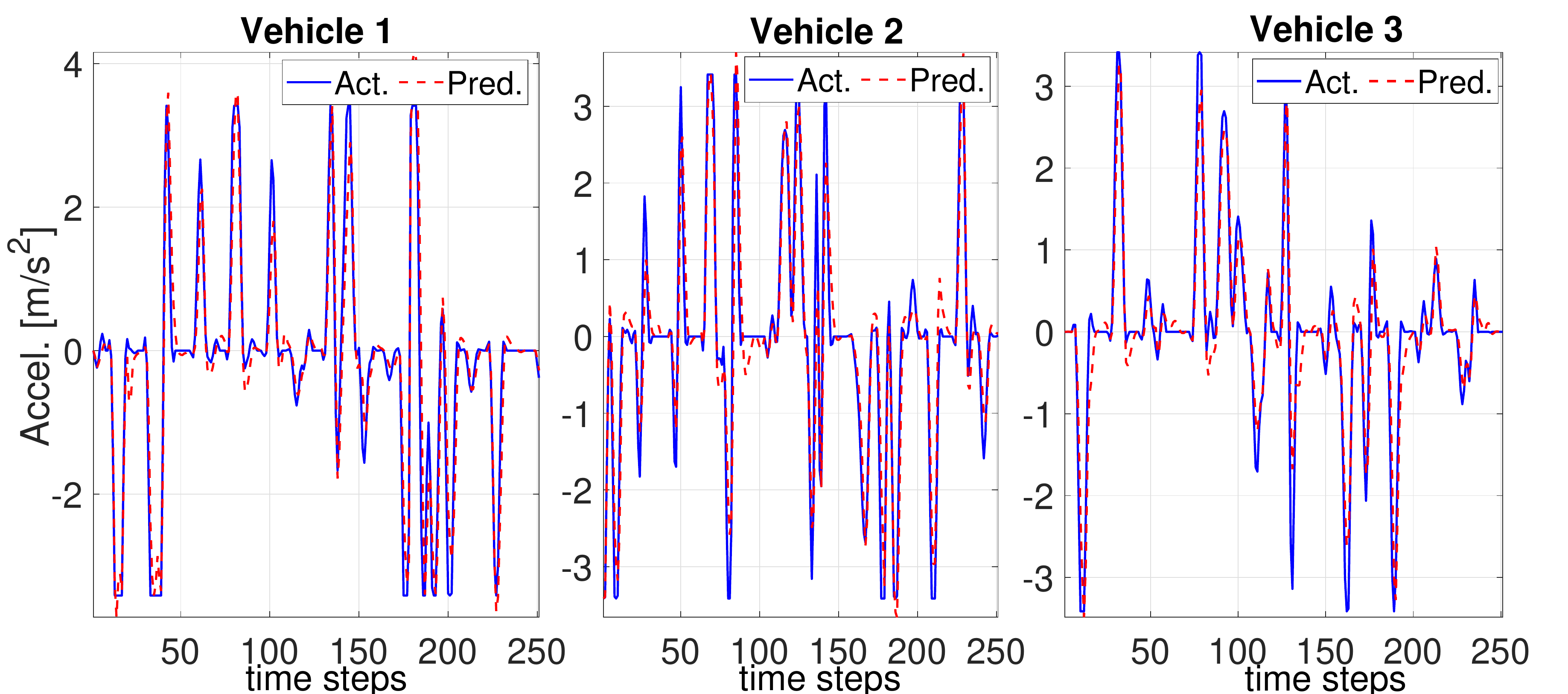}
 \caption{\small Comparison of the actual acceleration and the predicted acceleration using the real-time updated parameters for the first three vehicles in one NGSIM dataset.}\label{fig:Pred_NG}
\end{figure}
\begin{figure}[!h]
\centering
 \includegraphics[width=\linewidth]{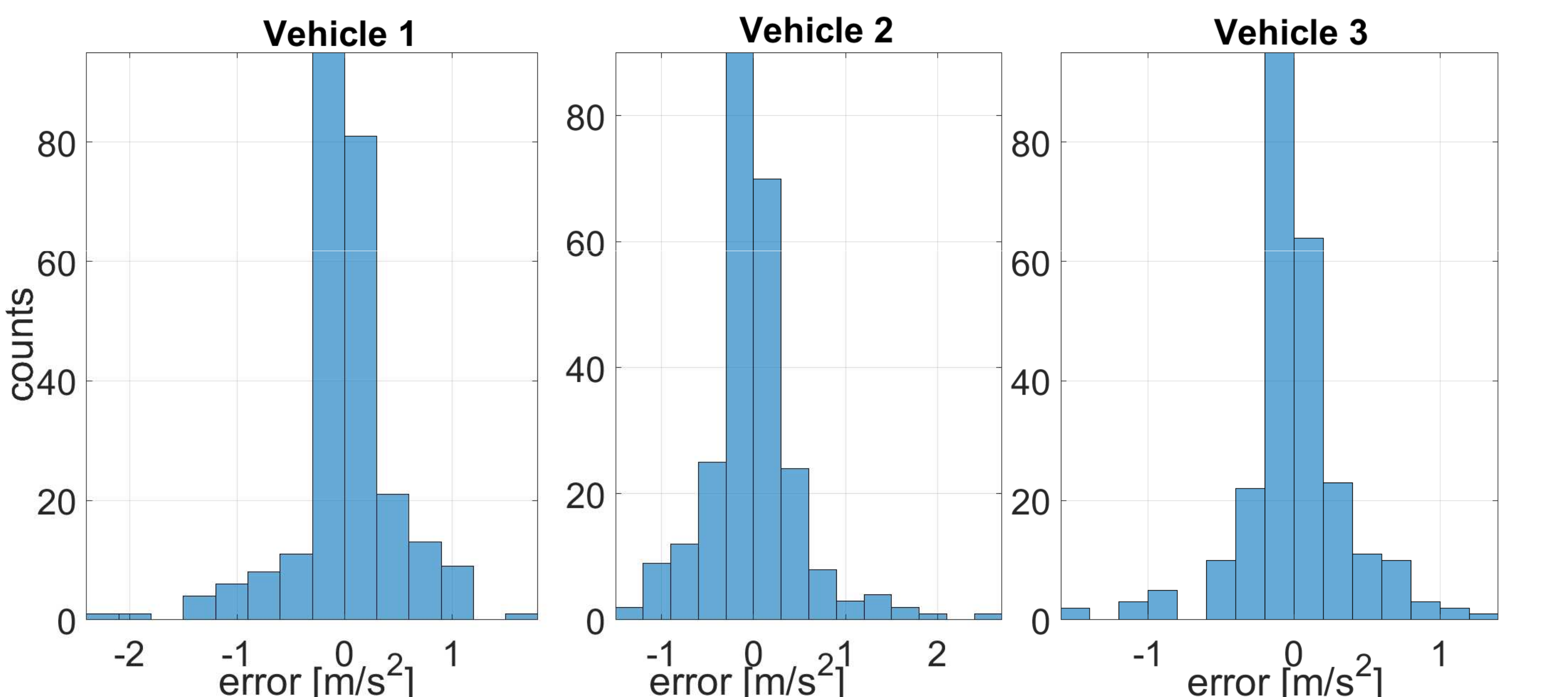}
 \caption{\small Prediction error distribution for the first three vehicles in one NGSIM dataset.}\label{fig:Error_NG}
\end{figure}

Furthermore, with the real-time identified parameters, the acceleration prediction is performed and compared with the measured one. We show this comparison and the prediction error distribution in Fig.~\ref{fig:Pred_NG} and Fig.~\ref{fig:Error_NG}, respectively, for  the first three vehicles. The other six vehicles have similar results and we omit here. The performance for all the 8 processed episodes is summarized in Table~\ref{tab:2}. It can be seen that the identified parameters are able to predict the driving trajectory (in terms of acceleration) with a good accuracy.

\begin{table}[!h]
\caption{Prediction performance summary on NGSIM}
\label{tab:2}
\centering
\begin{tabular}{c|c|c|c}
\hline
 Episode ID &\# of veh. &avg. RMSE [$m/s^2$]&worst RMSE [$m/s^2$] \\
 \hline
1 &9 &0.36 &0.39 \\
\hline
2 &6 &0.32 &0.45\\
\hline
3 &5 &0.40 &0.45 \\
\hline
4 &6 &0.26 &0.42 \\
\hline
5 &8 &0.33 &0.45\\
\hline
6 &7 &0.34 &0.45\\
\hline
7 &5 &0.41 &0.49\\
\hline
8 &5 &0.32 &0.47\\
\hline
\hline
\end{tabular}
\end{table}

\subsection{Naturalistic driving data from connected vehicles}
\begin{figure}[!h]
 \includegraphics[width=0.9\linewidth]{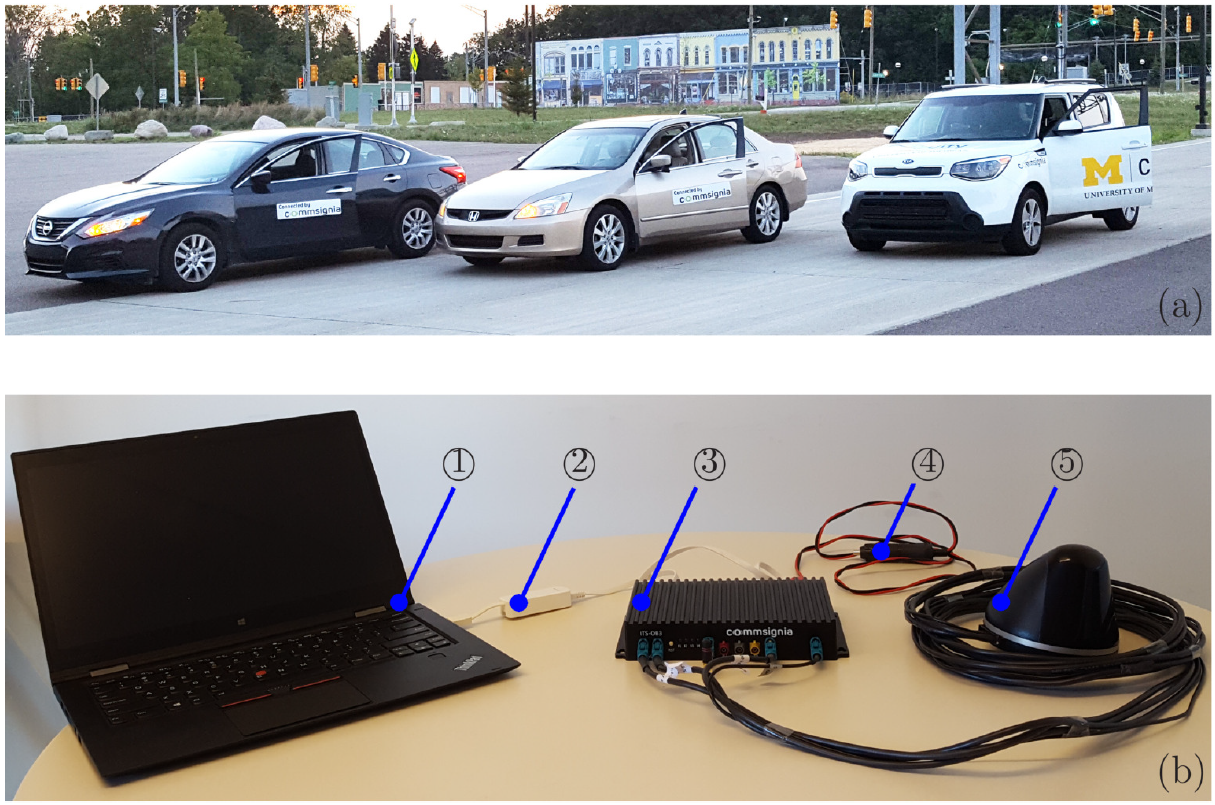}
 \caption{\small (a) Three connected vehicles used for data collection. (b) Vehicle-to-vehicle communication on-board unit: \textcircled{1} computer for data processing; \textcircled{2} Ethernet cable; \textcircled{3} electric control unit; \textcircled{4} power cable; and \textcircled{5} V2V/GPS antennae.}\label{fig:exp}
\end{figure}

We next evaluate the model and online parameter identification framework on the collected data from connected automated vehicle experiment in\cite{Exp_Setup}. In this setup, three vehicles are equipped and connected with dedicated short-range communication (DSRC) devices as shown in Fig.~\ref{fig:exp}(a). The position and speed data of each vehicle is recorded at a 10 Hz update rate. The real-time data of the vehicles are communicated and synchronized. 

\begin{figure}[!h]
\vspace{-8pt}
\centering
 \includegraphics[width=0.8\linewidth]{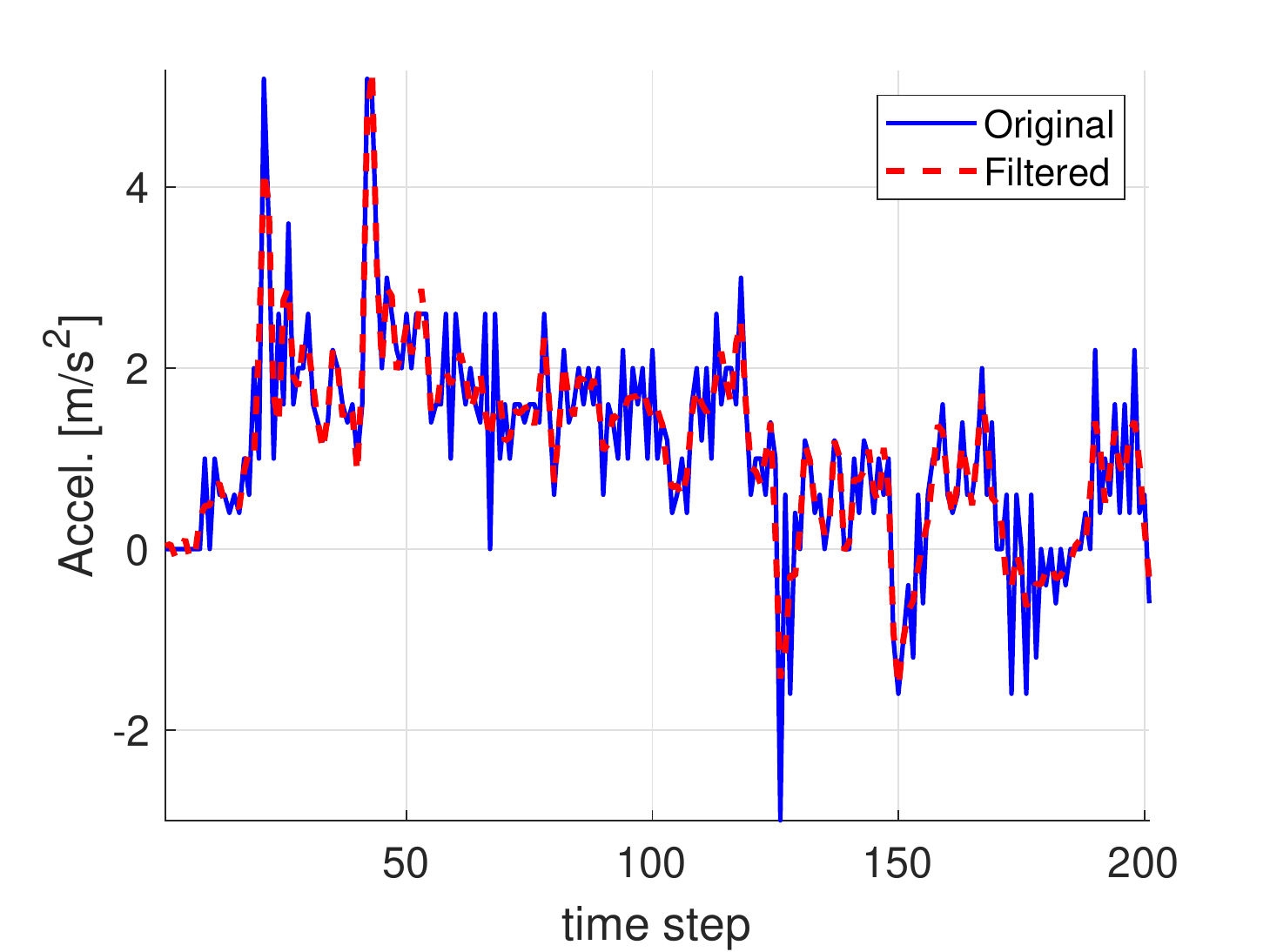}
 \caption{\small Low-pass filtering to remove high-frequency noises in acceleration data.}\label{fig:LPfilter}
\end{figure}

A total of four trips are recorded and as the collected acceleration measurements are noisy, we use a low-pass filter (Butterworth, 3 Hz cutoff frequency) to filter out high-frequency noises, which is illustrated in Fig.~\ref{fig:LPfilter}. 

\begin{figure}[!h]
\vspace{-5pt}
\centering
 \includegraphics[width=0.8\linewidth]{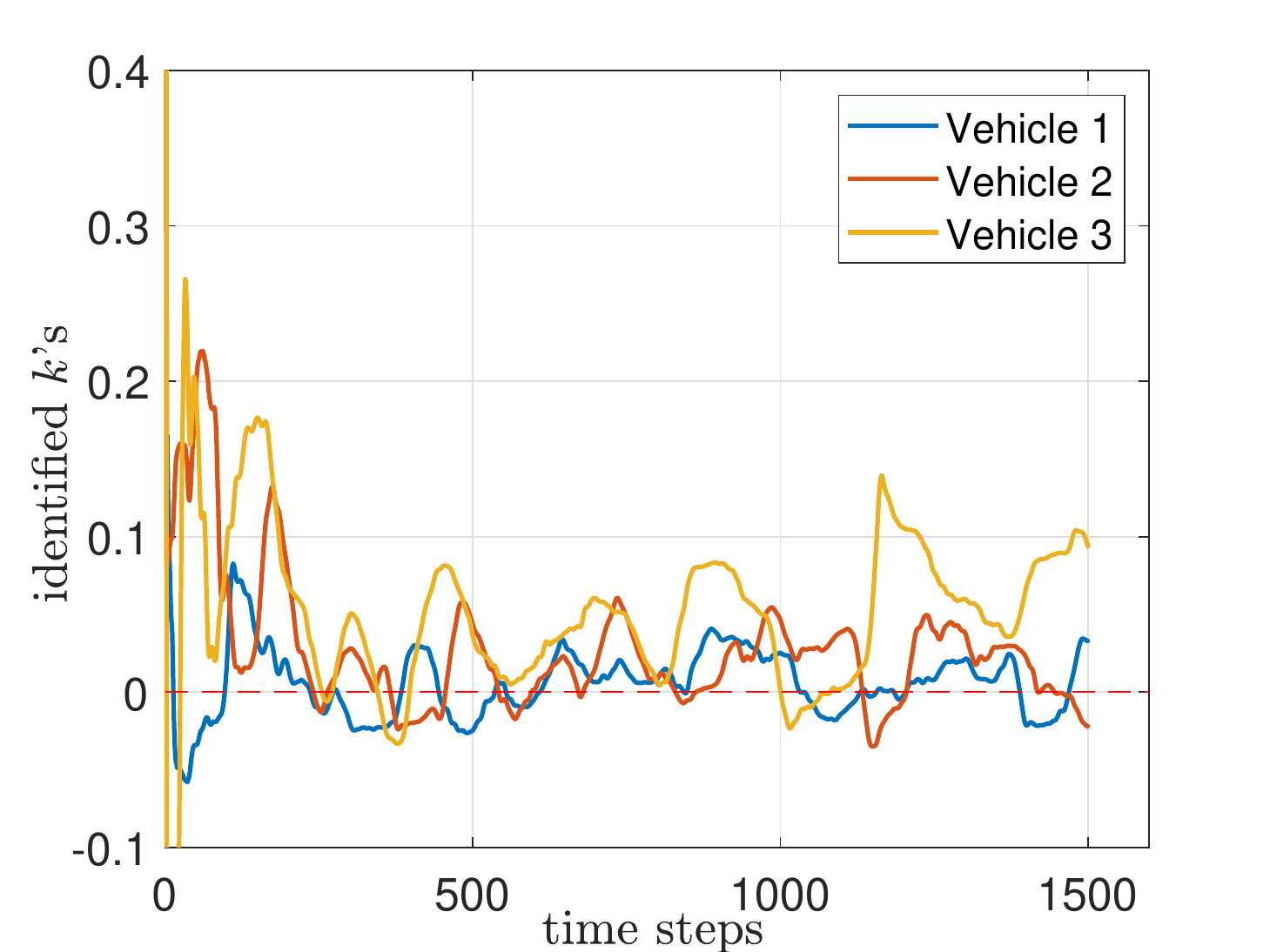}
 \caption{\small Real-time updated parameter $k$'s (spring stiffness)  for the 3 vehicles in trip 1.}\label{fig:paraKExp}
 \vspace{-5pt}
\end{figure}

\begin{figure}[!h]
\vspace{-15pt}
\centering
 \includegraphics[width=0.8\linewidth]{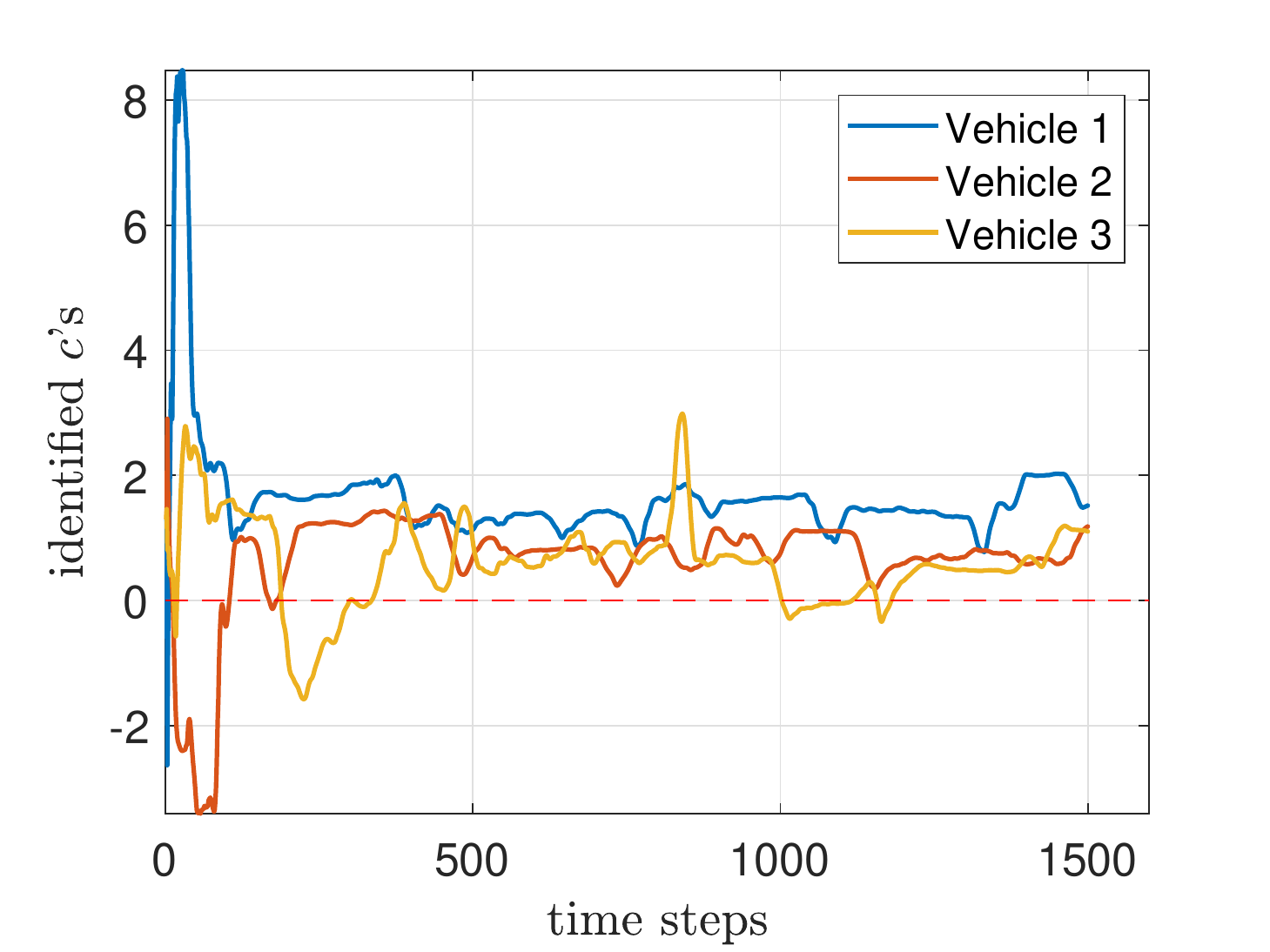}
 \caption{\small Real-time updated parameter $c$'s (damping coefficient)  for the 3 vehicles in trip 1.}\label{fig:paraCExp}
\end{figure}

We then run the SRLS-IQR algorithm on the processed data in an online fashion (i.e., feed the data step by step) with the hyper-parameters specified in Table~I. The identified parameters on the data of trip 1 are shown in Fig.~\ref{fig:paraKExp} and Fig.~\ref{fig:paraCExp}. Similar to the observations in NGSIM validation, we find that the $K$ and $C$ parameters are mostly stable and above 0. It is also observed that the parameters are more similar across the three drivers as compared to those in NGSIM. The reason might be that the vehicles were driven as a team effort so the drivers tend to observe other vehicles and drive similarly. 

\begin{figure}[!h]
\centering
 \includegraphics[width=\linewidth]{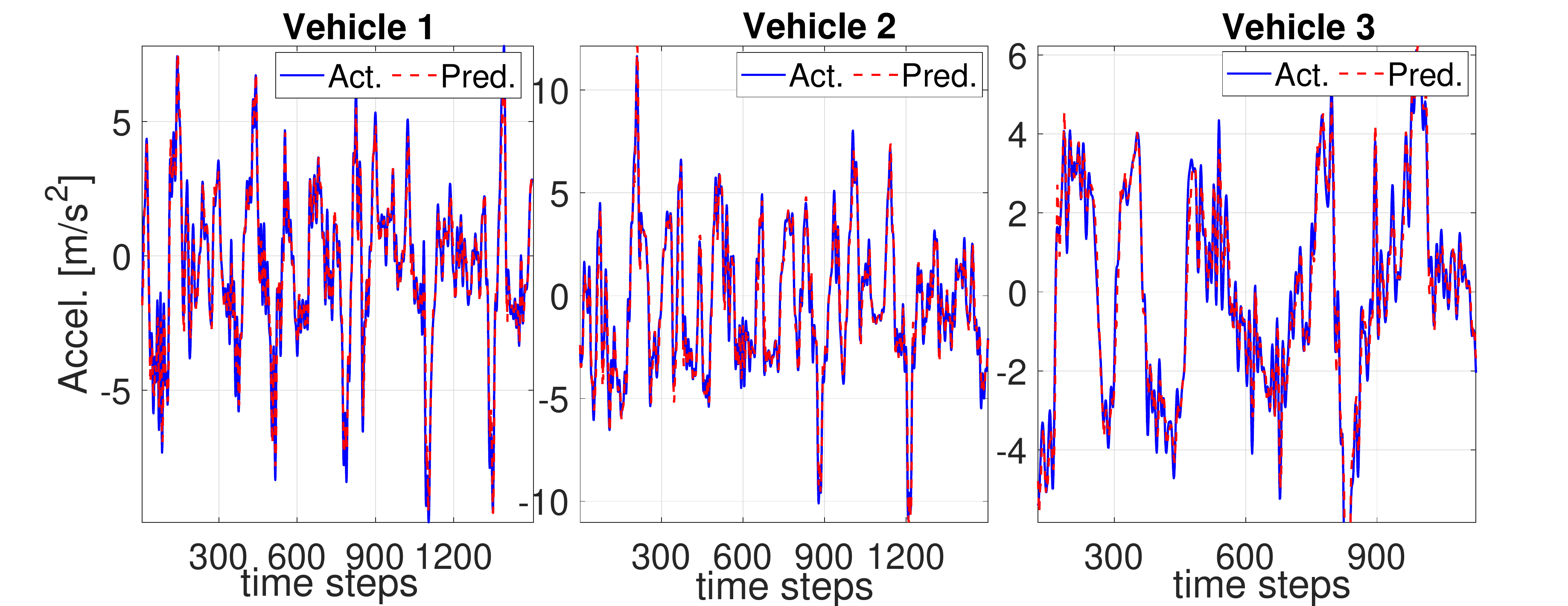}
 \caption{\small Comparison of the actual acceleration and the predicted acceleration using the real-time updated parameters on the data of trip 1.}\label{fig:Pred_Exp}
\end{figure}

\begin{figure}[!h]
\centering
 \includegraphics[width=\linewidth]{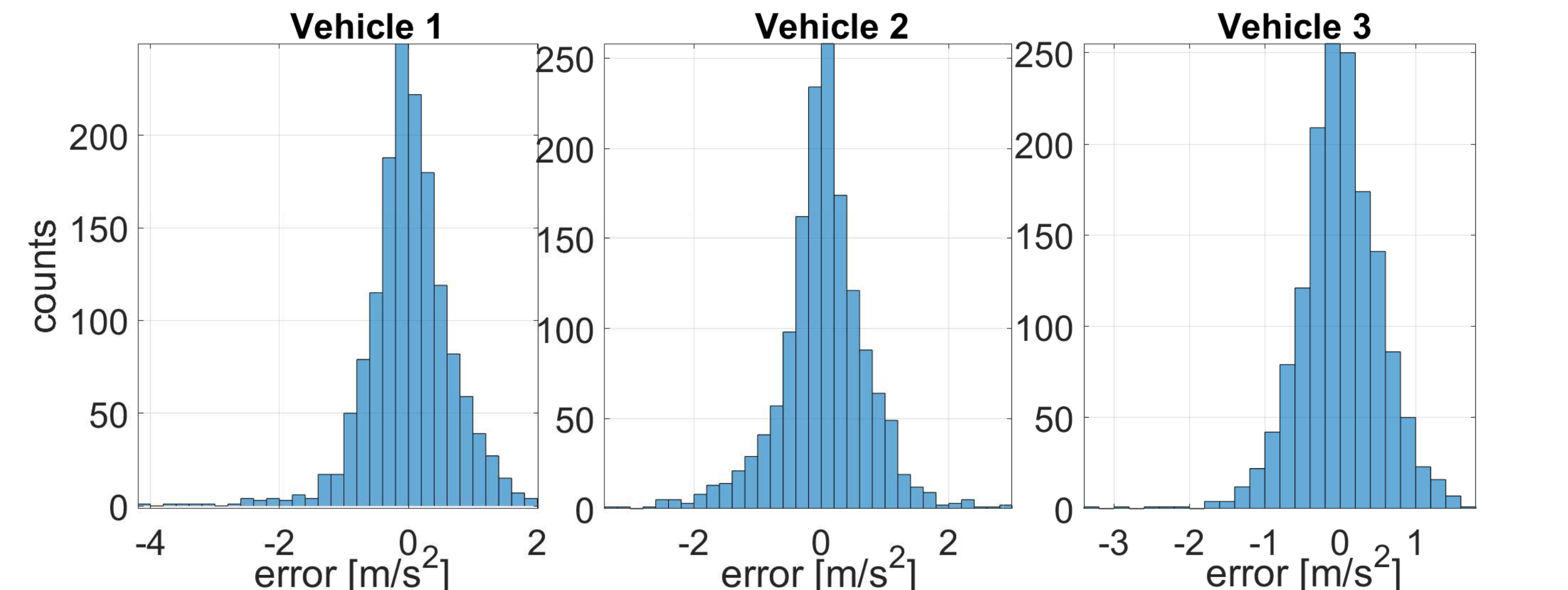}
 \caption{\small Prediction error distribution for trip 1.}\label{fig:Error_Exp}
\end{figure}

With the real-time identified parameters, the predicted and measured accelerations are shown in Fig.~\ref{fig:Pred_Exp} while the error distribution is shown in Fig.~\ref{fig:Error_Exp} for trip 1.  The performance for the all four  trips is summarized in Table~\ref{tab:3}. Overall, the identified parameters are able to predict the driving trajectory (in terms of acceleration) accurately. 

\begin{table}[!h]
\caption{\small Prediction performance summary on naturalistic driving data.}
\label{tab:3}
\centering
\begin{tabular}{c|c|c|c}
\hline
 Trip ID &\# of vehicles &avg. RMSE [$m/s^2$]&worst RMSE [$m/s^2$] \\
 \hline
1 &3 &0.28 &0.36 \\
\hline
2 &3 &0.22 &0.35\\
\hline
3 &3 &0.36 &0.44 \\
\hline
4 &3 &0.33 &0.39 \\
\hline

\hline
\end{tabular}
\end{table}

\section{Conclusions}
In this paper, we developed a new mechanical system inspired mass-spring-damper traffic model to characterize the car-following dynamics of a chain of vehicles. The model naturally captured human driving behaviors and considered the impact of the following vehicle on the lead vehicle. A sequential recursive least square with inverse QR decomposition algorithm was developed to identify the driving-related parameters in real time. The model and parameter identification framework were validated on two real-world driving datasets: NGSIM and our own collected vehicle data, and both demonstrated promising performance. 

In our future work, we plan to develop parameter identification algorithms that can identify heterogeneous scaling factors and time delays across the drivers. We will also consider the incorporation of nonlinear terms in the dynamics that can potentially improve the model performance. 

\section*{Acknowledgement}
The authors would like to thank Professor G\'{a}bor Orosz from the University of Michigan for his many insightful suggestions and comments.
\begin{IEEEbiography}
 [{\includegraphics[width=1in,height=1.25in,clip,keepaspectratio]{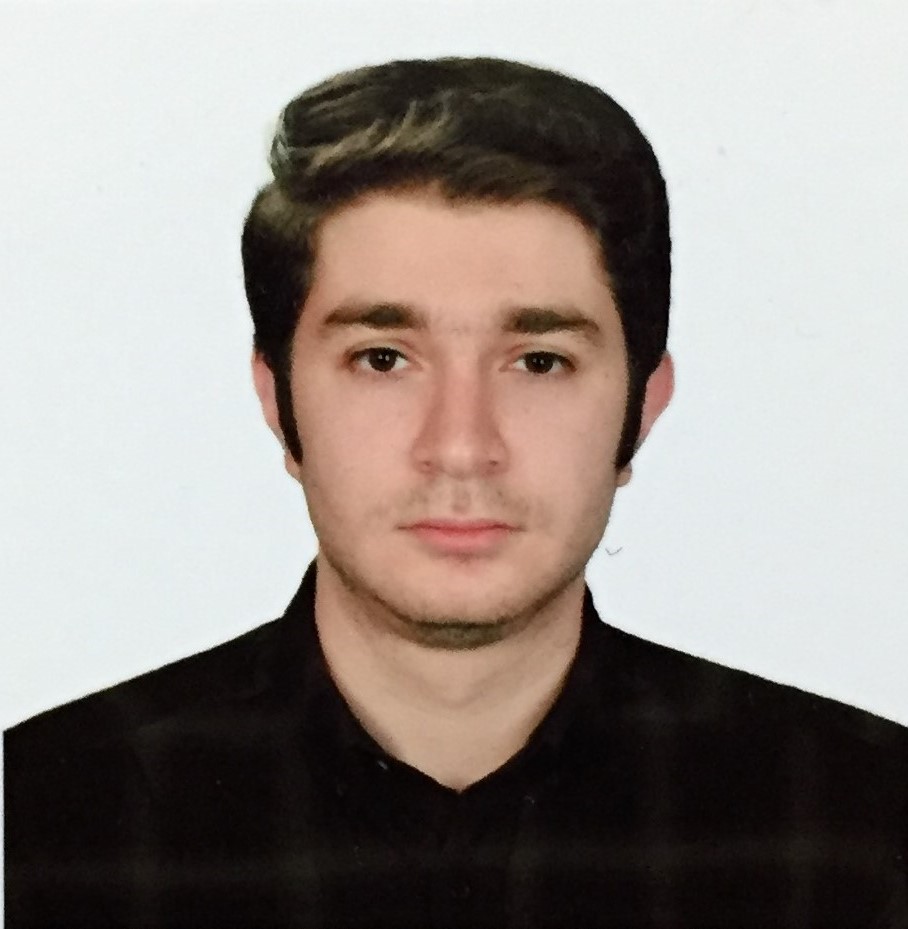}}]{Mohammad R. Hajidavalloo} obtained his B.Sc. and M.Sc. degree from  University of Tehran in Mechanical Engineering  in 2016 and 2018 respectively.

He is currently pursuing the Ph.D. degree in the department of Mechanical Engineering at Michigan State University. His research interests include Learning-based Control, Optimal Control and Reinforcement learning.
\end{IEEEbiography}
\begin{IEEEbiography}
 [{\includegraphics[width=1in,height=1.25in,clip,keepaspectratio]{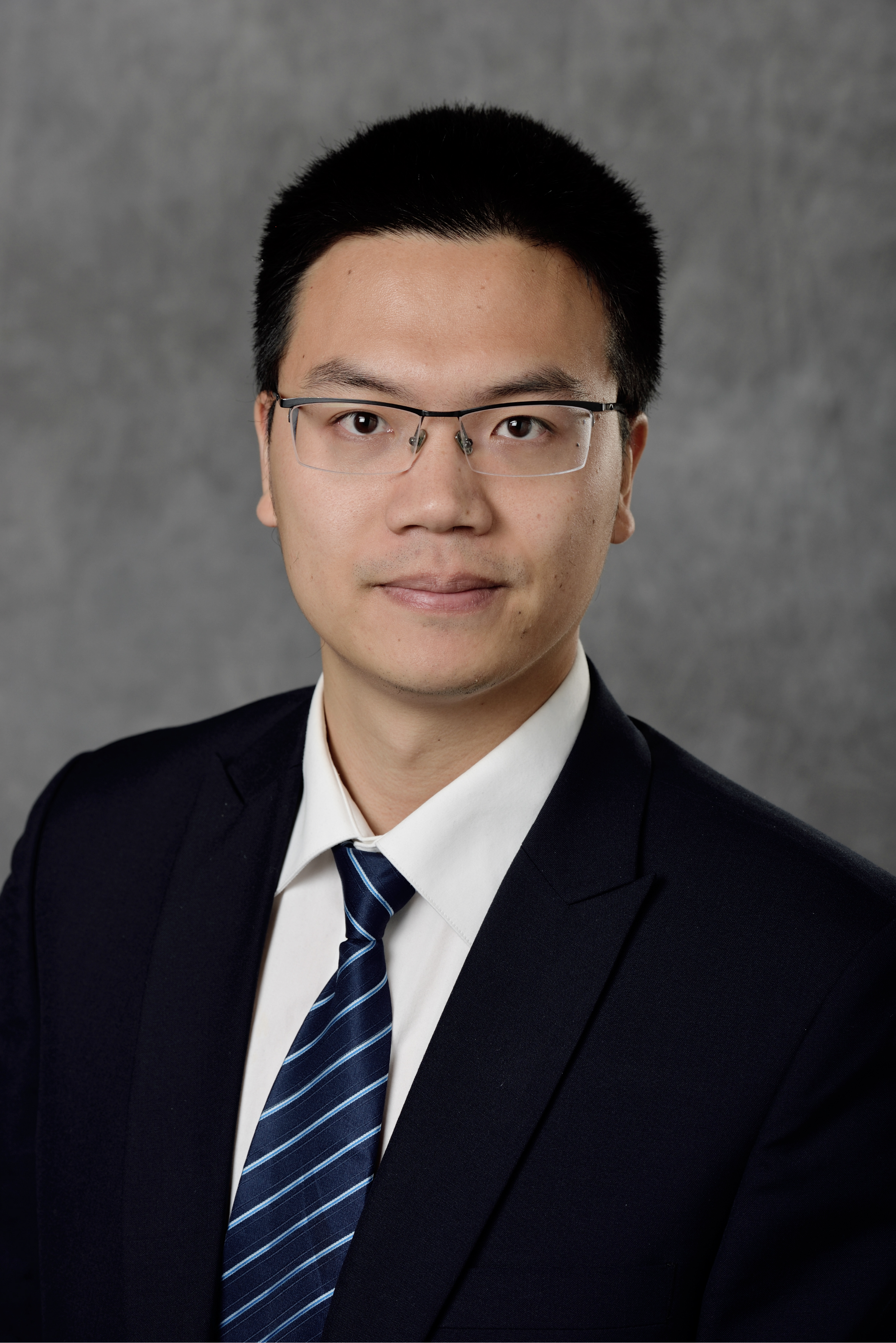}}]{Zhaojian Li} received his B. Eng. degree from Nanjing University of Aeronautics and Astronautics in 2010. He obtained M.S. (2013) and Ph.D. (2015) in Aerospace Engineering (flight dynamics and control) at the University of Michigan, Ann Arbor.

He is currently an Assistant Professor with the department of Mechanical Engineering at Michigan State University. His research interests include Learning-based Control, Nonlinear and Complex Systems, and Robotics and Automated Vehicles.
\end{IEEEbiography}
\begin{IEEEbiography}
 [{\includegraphics[width=1in,height=1.25in,clip,keepaspectratio]{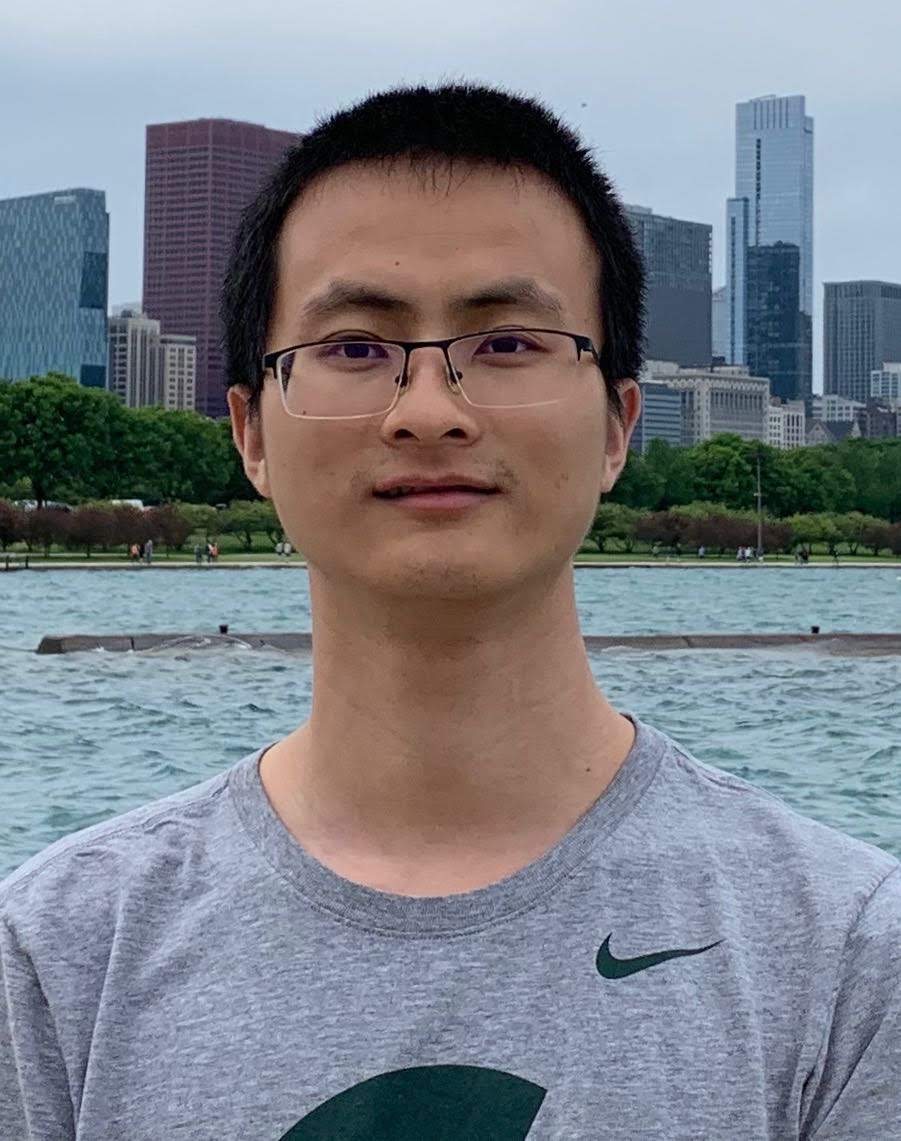}}]{Dong Chen}  received the B.E. degree from the UESTC, Sichuan, Chengdu, China, in 2017. He is currently pursuing the Ph.D. degree in the Department of Mechanical Engineering, Michigan State University, East Lansing, USA. His primary research interests include multi-agent systems, autonomous driving, and reinforcement learning.
\end{IEEEbiography}
\begin{IEEEbiography}
 [{\includegraphics[width=1in,height=1.25in,clip,keepaspectratio]{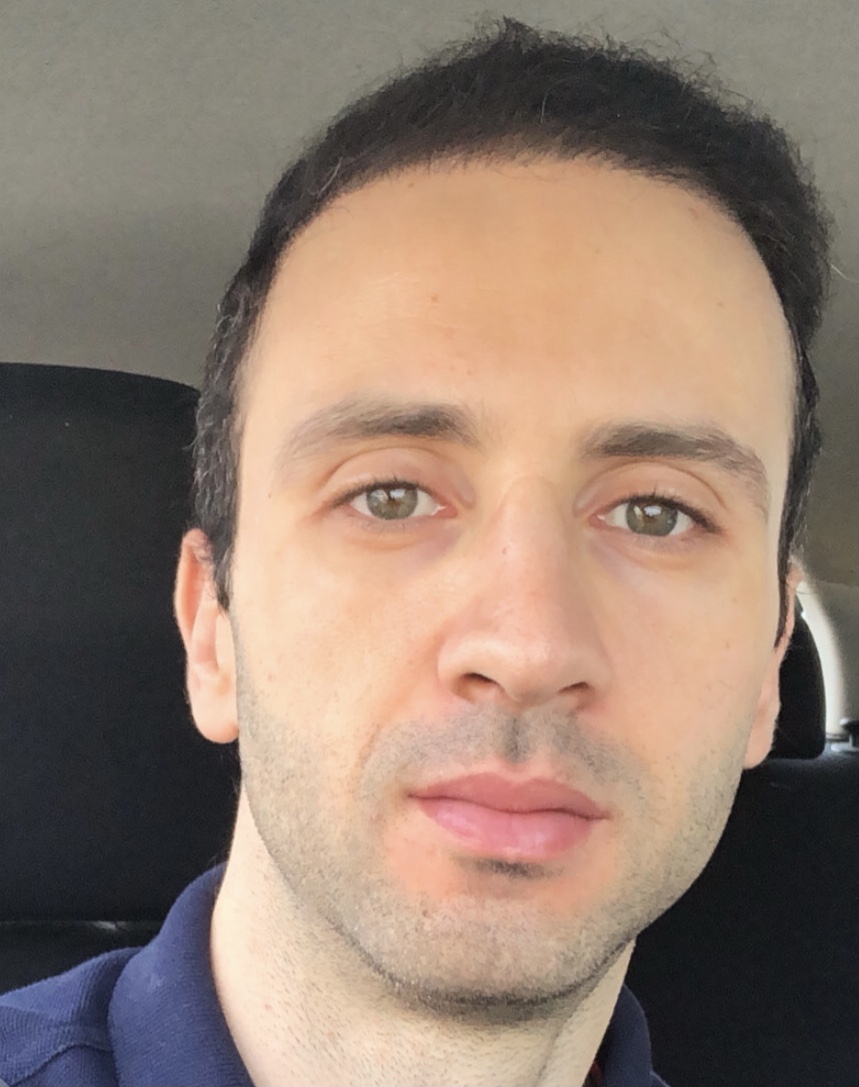}}]{Ali Louati} received the B.Sc. and M.Sc. degrees in computer science and business from the University of Sfax, Tunisia, and the Ph.D. degree in computer science and business from the ISG, University of Tunis, Tunisia, in 2010, 2013, and 2018, respectively. From 2014 until 2018, he was a research assistant at both, IFMA, LIMOS, Clermont-Ferrand, France, and King Saud University. Currently, he is an assistant professor at Prince Sattam Bin Abdulaziz University, Saudi Arabia. He is a member of the SMART Laboratory at ISG, University of Tunis. He has contributed to several research projects in Tunisia, France, and Saudi Arabia. His main research interests include machine learning, artificial immune systems, data mining, and multiobjective optimization. The main application domain consider by Ali Louati is the development of Intelligent Transportation Systems. He serves as a reviewer for several international journals such as European Transportation Research Review, the Journal of Supercomputing, and Applied Soft Computing.
\end{IEEEbiography}

\begin{IEEEbiography}
 [{\includegraphics[width=1in,height=1.25in,clip,keepaspectratio]{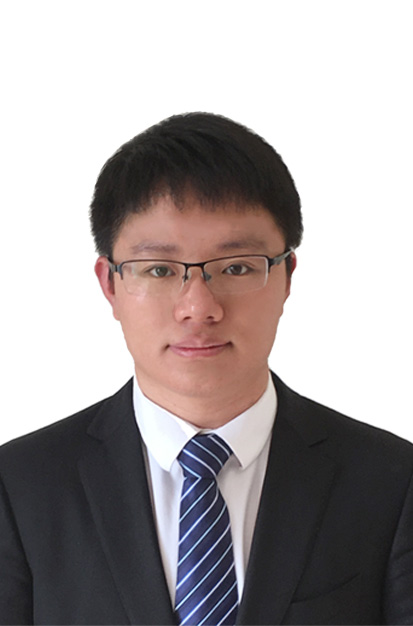}}]{Shou Feng} received the B.S. and Ph.D. degrees from the Department of Automation, Tsinghua University, China, in 2014 and 2019, respectively. He was also a visiting Ph.D. student in the Department of Civil and Environmental Engineering, University of Michigan, Ann Arbor, from 2017 to 2019, where he is currently a Post-Doctoral Researcher. He was the recipient of the IEEE ITSS Best Ph.D. Dissertation Award (2020). His current research interests include testing, evaluation, and optimization of connected and automated vehicles.
\end{IEEEbiography}
\begin{IEEEbiography}
 [{\includegraphics[width=1in,height=1.25in,clip,keepaspectratio]{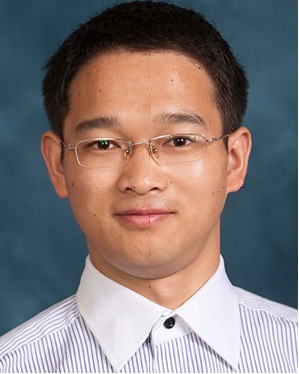}}]{Wubing B. Qin} received the B.Eng. degree from the School of Mechanical Science and Engineering, Huazhong University of Science and Technology, Wuhan, China, in 2011, and the M.Sc. and Ph.D. degree in mechanical engineering from the University of Michigan, Ann Arbor, MI, USA, in 2016 and 2018, respectively. Then he joined Aptiv, Michigan, USA, as an ADAS algorithm developer in 2018. Currently, he is a research engineer at Ford Motor Company since 2020. His research focuses on dynamics and control of connected automated vehicles, digital systems, ground robotics, and nonlinear and stochastic systems with time delays
\end{IEEEbiography}

\bibliographystyle{ieeetr}
\bibliography{main}

\end{document}